\documentclass[aps,groupedaddress,nofootinbib]{revtex4}

\usepackage{graphicx}
\usepackage{amssymb,amsmath}
\usepackage[usenames,dvipsnames]{color}

\newcommand{\be}{\begin{equation}}
\newcommand{\ee}{\end{equation}}
\newcommand{\bea}{\begin{eqnarray}}
\newcommand{\eea}{\end{eqnarray}}	

\newcommand{\ba}{\begin{array}}
\newcommand{\ea}{\end{array}}

\definecolor{MyDarkGray}{RGB}{140,140,140}

\begin{document}
\begin{flushleft}
KCL-PH-TH/2011-25 \\
LCTS/2011-08 \\
CERN-PH-TH/2011-191\\
\end{flushleft}

\title{On the R\^ole of Space-Time Foam in Breaking Supersymmetry via the Barbero-Immirzi Parameter}



\author{John Ellis and Nick E. Mavromatos}
\affiliation{Theoretical Particle Physics and Cosmology Group, Department of Physics, King's College London, Strand, London WC2R 2LS, UK; \\
Theory Division, Physics Department, CERN, CH-1211  Geneva 23, Switzerland}



\begin{abstract} 

We discuss how: (i) a dilaton/axion superfield can play the
r\^ole of a Barbero-Immirzi field in four-dimensional conformal quantum supergravity theories,
(ii) a fermionic component of such a dilaton/axion superfield may play the r\^ole of a Goldstino in
the low-energy effective action obtained from a superstring
 theory with $F$-type global supersymmetry breaking,
 (iii) this global supersymmetry breaking is communicated to the gravitational sector
via the supergravity coupling of the Goldstino, and
(iv) such a scenario may be realized explicitly in a D-foam model
with D-particle defects fluctuating stochastically.
\end{abstract}

\maketitle

\section{Introduction}

One of the most important issues in string theory and supersymmetry phenomenology
is the mechanism of supersymmetry breaking. It is generally thought to be
non-perturbative, but developing the intuition and calculational
tools needed to understand the relevant aspects of non-perturbative string theory is
an unmet challenge, as yet.
Given these shortcomings in our understanding, it is natural to reason by
analogy with better-understood non-perturbative phenomena in field theory,
notably in gauge theories and specifically in (supersymmetric) QCD.
Much intuition has been obtained from the well-studied phenomenology
of non-perturbative effects in QCD, which has been bolstered by explicit
exact calculations in $N = 1$ and $N = 2$ supersymmetric gauge theories.
Building on this experience, promising scenarios have been proposed for
supersymmetry breaking via gaugino condensation induced by
non-perturbative gauge dynamics.
We have no objection to such scenarios, but cannot resist trying to be
more ambitious, and develop a scenario for supersymmetry breaking
rooted in intrinsically non-perturbative string dynamics.

From this point of view, it is natural to consider intrinsically stringy
analogues of non-perturbative aspects of gauge theories. The latter
have topologically non-trivial sectors populated by gauge configurations
such as instantons and monopoles, whose contributions to the gauge
functional integral are weighted by arbitrary vacuum angle parameters
such as $\theta_{QCD}$. In the case of string theory, there are various
classes of non-perturbative D-brane configurations and analogues of
gauge vacuum angle parameters. The questions then arise how they
might lead to some condensation phenomenon, analogous to quark
or gluino condensation, that might break supersymmetry.

In this paper, our approach to these questions takes as its starting-point
the effective Lagrangian that characterizes the low-energy, large-distance
limit of string theory. This should be taken to have the most general form
compatible with basic symmetry principles, which we take to be $N = 1$
local supersymmetry, i.e., supergravity. As we discuss below,
this includes the supersymmetric extension of the Holst term proportional to the
dual of the Riemann tensor, as well as the conventional Einstein action. The coefficient
of this term, called the Barbero-Immirzi parameter $\gamma$,
has some superficial affinity with a $\theta$ term in a non-Abelian gauge theory.
The supersymmetric completion of the Host term involves a dilaton/axion/dilatino
supermultiplet, and we explore whether this could play a r\^ole in local
supersymmetry breaking, with the dilatino becoming the Goldstino.

For this to happen, there should presumably be some stringy analogue
of the instanton and related non-perturbative non-Abelian gauge theory
configurations that play a key r\^ole in quark condensation and chiral
symmetry breaking in QCD, and gluino condensation in $N \ge 2$
supersymmetric gauge theories. As discussed elsewhere,
D-particles moving in the higher-dimensional bulk appear in a 3-dimensional
brane world, such as the one we inhabit, to be localized at space-time events
$x_\mu$, just like instantons. In order to develop the analogy further, one
should identify the dynamical interaction of matter with D-particles that
might give rise to condensation. We have argued that fermions without
internal quantum numbers such as the dilatino may indeed have non-trivial
interactions with D-particles, and in this paper we sketch how such interactions
might play a r\^ole in condensation and supersymmetry breaking.
We regard this is an illustrative example how short-distance Planckian
dynamics might play a r\^ole in supersymmetry breaking.

\section{Quantum Gravity, the Barbero-Immirzi Parameter and Supergravity}

\subsection{Introducing the Barbero-Immirzi Parameter}

The Ashtekar formalism of General Relativity (GR)~\cite{ashtekar} is a version of canonical quantisation, which
introduces self-dual SL(2,C) connections as the fundamental underlying variables, enabling the General Relativity (GR)
constraints to be reduced to a polynomial form. However, the complex nature of the self-dual connections 
necessitates the introduction of reality conditions, which complicate the quantisation procedure. For the moment, 
this hurdle prevents the emergence of a complete theory of Quantum Gravity
from such a procedure. In order to avoid this problem, Ashtekar~\cite{asht2} and Barbero~\cite{barbero} independently 
introduced real-valued SU(2) or SO(3) connections (called Ashtekar-Barbero connections) in a partially `gauge fixed''
vierbein formalism of GR. This led Ashtekar, Rovelli and  Smolin subsequently to the development of the Loop Quantum 
Gravity (LQG) approach~\cite{loopqg} to the quantisation of GR, according to which one can construct a non-perturbative 
and space-time background-independent formalism for QG, which in some explicit examples is free from the short-distance 
singular behaviour of GR.

The Ashtekar-Barbero connections contain a free parameter, $\gamma$, which arises when one expresses the Lorentz 
connection of the non-compact group SO(3,1) in terms of a complex connection in the compact groups of SU(2) or SO(3). 
The existence of such a free parameter in the connection of the Ashtekar formalism was implicit in the formalism of 
Barbero~\cite{barbero}, but was put in a firm footing by Immirzi~\cite{immirzi}, and is now
termed the Barbero-Immirzi parameter. The significance of this parameter became obvious after the observation that the 
area operator of LQG depends on it, leading to a black-hole entropy in this formalism of the form 
(in four-dimensional Planck units $M_P=1$): $S = \frac{\gamma_0}{\gamma} \frac{A}{4} $, where 
$\gamma_0 $ is a numerical factor depending on the gauge group. The Standard Bekenstein-Hawking  entropy is 
recovered in the case $\gamma_0=\gamma$.

\subsection{ Fermionic Torsion and the Barbero-Immirzi Parameter}

Fermions induce a non-trivial torsion term into the gravitational action that
involves, in the first-order formalism, the dual of the Riemann curvature form.
This introduces a new parameter in the action, namely the  Barbero-Immirzi parameter~\cite{immirzi}, $\gamma$, introduced above.
Specifically, using the Palatini formalism of general relativity to express the
four-dimensional Einstein-Hilbert action in terms of the vierbeins
$e^\mu_m $ and the spin connection,  $\omega_{\mu}^{m n}$~\footnote{We remind the reader that Latin indices and 
quantities with a tilde refer to the flat Minkowski tangent space-time plane.},
as is necessary in the presence of fermions, one can always add to the action
a term involving the dual of the curvature tensor,
$^\star R^{m n}_{\,\,\,\,\,\,\,\mu\nu} \equiv \frac{1}{2}\epsilon^{m n}_{\,\, \,\,\, p q}R^{p q}_{\,\,\,\,\,\mu\nu}$, obtaining~\cite{holst}:
\bea\label{gravaction}
S_{\rm grav} (e, \, \omega) &=& S_{\rm Einst} + S_{\rm Host}~, \nonumber \\
{\rm where} \; \; \; \; \;
S_{\rm Einst} &=& \frac{1}{16\pi G_N} \int d^4 x e \, e^\mu_{\,\, m} \, e^\nu_{\,\, n} R^{m n}_{\,\,\,\,\,\mu\nu}~, \nonumber \\
{\rm and} \; \; \; \; \; \;
S_{\rm Holst} &=& - \frac{1}{2\,\gamma}\frac{1}{16\pi G_N} \int d^4 x e \, e^\mu_{\,\, m} \, e^\nu_{\,\, n} \epsilon^{m n}_{\,\,\,\,\, p\, q}R^{p q}_{\,\,\,\,\,\mu\nu}~.
\eea
Viewing gravity as a gauge theory, the second term (the Holst modification to the Einstein action)
has an arbitrary coefficient $1/\gamma$ that is somewhat analogous to the $\theta$
parameter in a non-Abelian gauge theory. The presence of the second term, if it is non-trivial,
induces an antisymmetric term in the connection, i.e., non-trivial torsion:
\begin{equation}\label{connx}
\omega_\mu^{a b} = {\tilde \omega}_\mu^{a b} + C_\mu^{a b}~,
\end{equation}
where ${\tilde \omega}^{ab}$ denotes the torsion-free connection determined by the tetrads
and $C_\mu^{a b}$ is the torsion. In the limit $\gamma \rightarrow 0$, the torsion term vanishes in a
path integral over the Euclidean gravity action, and this corresponds to the standard torsion-free limit
of general relativity.

In pure gravity, the Holst term does not affect the graviton equation of motion, which takes the form:
\begin{equation}\label{zerotorsion}
D_{\left[\mu \right.} e^m_{\left.{\nu}\right]} = 0~,
\end{equation}
where
\begin{equation}\label{covariant}
D_{\mu} = \partial_\mu  - \frac{i}{4} \omega_\mu^{a\,b}\sigma_{ab}~,
\end{equation}
with $\sigma^{ab} \equiv \frac{i}{2}\left[{\tilde \gamma}^a, {\tilde \gamma}^b\right]$, denotes the
gravitational covariant derivative.
In view of the Bianchi identity, $R_{\left[ \mu\nu\rho \right]\sigma} = 0$, (\ref{zerotorsion}) implies that the
torsion (Barbero-Immirzi) term in (\ref{gravaction}) is identically zero in pure general relativity in the absence of matter.
In such a case, only the $\tilde{\omega}_\mu^{ab}$ term survives in the connection (\ref{connx}).

This is no longer the case when fermions $\psi$ are present~\cite{rovelli},
in which case it can be shown that the torsion term is non-zero.
This is because (\ref{connx}) is no longer satisfied, but one has instead
a non-vanishing right-hand side, as a result of non-zero fermionic currents.
In this case, the gravitational action is augmented by a fermion contribution:
\bea\label{gravaction2}
S_{GF} &=& S_{\rm grav} + \frac{i}{2} \int d^4 x \overline{\psi} \left(\gamma^\mu D_\mu (\omega) \psi -
\overline{ D_\mu (\omega) \psi} \gamma^\mu \psi \right)~.
\eea
Variation of the action (\ref{gravaction2}) with respect the connection $\omega_\mu^{ab}$ (\ref{connx}),
which incorporates torsion, yields:
\bea\label{defermi}
&& D_\mu \left( e \, e^\mu_{\left[ a \right.} e^\nu _{\left. b \right]} \right)= 8 \pi {p^{-1}_{a b}}^{\,\,\, c d} J_{c d}^\nu ~, \nonumber \\
&& {p^{-1}}^{a b}_{\,\,\, c d} = \frac{\gamma^2}{\gamma^2 + 1} \left(\delta^a_{\left[c \right.}\delta^b_{\left. d\right]}
+ \frac{1}{2 \gamma} \epsilon_{cd}^{\,\,\, a b} \right)~, \quad
J_{c d}^\mu = \frac{1}{4} e^\mu_a \, \epsilon^a_{c d f} j_{(A)}^f ~,
\eea
where $j_{(A)}^f \equiv \overline{\psi} \gamma_5 {\tilde \gamma}^f \psi $ is the fermionic axial current.
The torsion can be found in this case as a consistent solution of (\ref{defermi}),
and takes the form~\cite{rovelli}:
\begin{equation}\label{torsionf}
C_\mu^{a b} = - 2\pi G_N \frac{\gamma}{\gamma^2 + 1} \left( e^{\left[ a \right.}_\mu \, J_{(A)}^{\left. b \right]} -
\gamma e^c_\mu {\epsilon_{c}^{\,\,ab}}_{d} J_{(A)}^d \right) ~, \quad J_{(A)}^a = \overline{\psi} {\tilde \gamma}^a {\tilde \gamma}_5 \psi ~.
\end{equation}
From a path-integral point of view, the use of the equations of motion is equivalent to integrating
out the (non-propagating) torsion field.

It is straightforward to show in this approach~\cite{rovelli} that the effective Dirac
action contains an axial current-current interaction, with a coefficient that depends on the
Barbero-Immirzi parameter. To this end, consider the Dirac action coupled to a
connection field $\omega$ with torsion. Using the explicit form (\ref{covariant}) of
the covariant derivative, and the following property of the product of three flat-space
$\gamma$ matrices in four space-time dimensions:
\begin{equation}
{\tilde \gamma}^a {\tilde \gamma}^b {\tilde \gamma}^c = \eta^{a b} {\tilde \gamma}^c  + \eta^{b c} {\tilde \gamma}^a -
\eta^{a c} {\tilde \gamma}^b - i \epsilon^{a b c d }{\tilde \gamma}_5 {\tilde \gamma}_d ~,
\label{pro3gamma}
\end{equation}
we may write the fermionic action as follows~\cite{mukho}:
\begin{equation}\label{magfield}
\mathcal{L} = e \overline{\psi} \left(i {\tilde \gamma}^a \partial_a - m + {\tilde \gamma}^a {\tilde \gamma}^5 B_a \right) \psi~,
\quad B^a = \epsilon^{a b c d } \left(e_{b \lambda} \partial_a e^\lambda_c + C_{b c d} \right)~,
\end{equation}
where $C_{ b c d}$ denotes the torsion part of the connection (\ref{connx}),
which cannot be expressed in terms of the vierbeins (tetrads).
We see that in this formalism the field $B_a$ plays the r\^ole of an axial `external field':
its spatial components $\vec{B}$ act as a `magnetic' field,
while its temporal component
behaves as an axial `scalar potential' $B^0$. Even in flat space-times, the field $B_\mu$ is non trivial
in the presence of fermionic torsion.

Substituting the expression (\ref{torsionf}) for the torsion into (\ref{magfield}),
and using that $\epsilon_{a i_1 i_2 i_3} \epsilon^{b i_1 i_2 i_3} = 3 ! \delta_a^b$, we
straightforwardly arrive at an effective four-fermion Thirring-type interaction
that is quadratic in the axial fermion current $J_{(A)}^\mu$, of the form~\cite{rovelli}:
\begin{equation}\label{fourfermi}
S_{\rm int}  = - \int d^4 x \, e \, \frac{3}{2} \pi G_N \left( \frac{\gamma^2}{\gamma^2 + 1} \right) \,J_{(A)}^2~: \quad J_{(A)}^a = \overline{\psi} {\tilde \gamma}^a {\tilde \gamma}_5 \psi~,
\end{equation}
where $e = \sqrt{{\rm det}(g)} $ is the vierbein determinant. We note that the only remnant of the metric in this
four-fermi interaction is this vierbein determinant factor, as the rest of the terms
can be expressed in terms of flat space-time quantities alone, as a result of the properties of the vierbein.

We note that, when taking the complex conjugate of the action, in other words when the
covariant derivatives in the Dirac equation are taken to act on both the fermionic fields  $\psi$
and their conjugates $\overline{\psi}$, the contributions of only one part of the torsion (\ref{torsionf})
survive in the Hermitean effective action (\ref{fourfermi}), namely that proportional to
$\gamma^2/(\gamma^2 + 1)$ and the dual of the axial current ${\epsilon_c^{\,\, a b}}_d j_{(A)}^d $.
This allows the Barbero-Immirzi parameter to assume purely imaginary values, and thus yield
\emph{attractive} four-fermion interactions. Such interactions might then lead to dynamical
mass generation for  the fermions, and thus chiral symmetry breaking in the case of multiflavour
interactions~\cite{neubert}~%
\footnote{Cosmological aspects of such an approach, in which the dynamical
fermion condensate may be identified with the dark energy at late eras of the Universe,
have been discussed in~\cite{biswas}.}.

However, the appearance of the Barbero-Immirzi parameter in the effective action
in the above approach~\cite{rovelli}, as a four-fermion interaction coupling, would invalidate the analogy
of the Holst action with the instanton action of QCD, in which the $\theta$ angle is \emph{purely topological}~\footnote{We
note, however, that instantons are argued to lead to effective chiral-symmetry-breaking multi-fermion interactions.}.
Moreover, the loop-quantum-gravity limit, in which $\gamma \to \pm i$, would lead to divergent
four-fermion couplings (\ref{fourfermi}), incompatible with the well-defined Ashtekar-Romano-Tate theory~\cite{art},
a version of the canonical formulation of quantum gravity in which only the self-dual parts of the curvature tensor
contribute to the Lagrangian density.

The above approach has been criticized~\cite{mercuri} on the grounds of mathematical inconsistency,
namely that, for an arbitrary value of the Immirzi parameter, the decomposition of the torsion (\ref{torsionf}) into
its irreducible parts, a trace vector, a pseudo-scalar axial vector and a tensor part, fails for the following reason.
Consider the contorsion tensor
\bea
C_\mu^{a b} = C^{\nu \rho}_\mu e^a_{\left[\nu\right.}e^b_{\left. \rho\right]}~,
\eea
and decompose it into its irreducible parts, namely the trace vector $C_\mu = C^{\nu}_{\mu \nu}$,
the pseudo-trace axial vector $S_\mu = \epsilon_{\mu\nu\rho\sigma} C^{\nu\rho\sigma}$ and the
tensor $q_{\mu\nu\rho}$ (with $q^\nu_{\rho \nu} = 0, \, \epsilon_{\mu\nu\rho\sigma}q^{\nu\rho\sigma} = 0$).
The solution (\ref{torsionf}) would imply~\cite{mercuri}
\bea\label{trace}
T^\mu = \frac{3}{4} \frac{\gamma}{\gamma^2 + 1} J_{(A)}^\mu ~, \quad
S^\mu = - \frac{3 \gamma^2}{\gamma^2 + 1} J_{(A)}^\mu~, \quad q^{\mu\nu\rho} = 0~.
\eea
The first of these relations is inconsistent, as it equates a Lorentz vector ($T^\mu$) with a pseudovector ($J_{(A)}^\mu$),
which have different transformation properties under the Lorentz group. As such, in this formulation, the Barbero-Immirzi
parameter cannot be arbitrary: the only consistent limiting values are
\bea\label{limits}
 {\rm either}  \quad &\gamma & \rightarrow 0~ \qquad~, \nonumber \\
 {\rm or} \quad &\gamma & \rightarrow \infty ~ \quad ~.
\eea
In the first limit there is no torsion at all, and in the second limit the torsion
is given by the fermionic axial vector, which is a result characteristic of the Einstein-Cartan theory:
\begin{equation}\label{torsionec}
C_{\mu a b} = - \frac{1}{4} e^c_\mu \epsilon_{a b c d} \overline{\psi} \gamma_5 {\tilde \gamma}^d \psi ~.
\end{equation}
In both the limits (\ref{limits}), the trace $C_\mu$ of the contorsion tensor vanishes, and thus the theory is consistent.

In fact, once the torsion assumes the Einstein-Cartan form (\ref{torsionec}), it is straightforward to show,
using (\ref{magfield}), that the effective Dirac action contains an axial current-current interaction, with a fixed coefficient, independent of the Barbero-Immirzi parameter:
\begin{equation}\label{intec}
S_{\rm int}^{EC} = - \int d^4 x e \frac{3}{2} \pi G J_{(A)}^a J_{(A)\, a}~.
\end{equation}
This is the limiting case of (\ref{fourfermi}) when $\gamma \rightarrow + \infty$,
but with $\gamma$ a \emph{real} parameter. However, we stress once more,
attractive four-fermion interactions arise in the
approach of \cite{rovelli} only in the case of a \emph{purely imaginary} Barbero-Immirzi parameter~\cite{rovelli}.
Moreover, as we see below, this limit respects local supersymmetry transformations.

To avoid the above-mentioned constraint on the Immirzi parameter $\gamma$,
and thus incorporate in a consistent way the limit $\gamma = \pm i$,
non-minimal couplings of the Holst action to fermions were considered in~\cite{mercuri,alexandrov,mercuri2},
that allow for arbitrary values of the Barbero-Immirzi parameter. In this way
the inconsistency in (\ref{trace}) is removed and the analogy of this parameter with
the $\theta$ angle of QCD is more complete.
Specifically, one may consider a non-minimal fermion coupling in the Holst action
\begin{equation}
S_{\rm Holst} = \frac{i \eta}{2} \int d^4 x e \left[ \overline{\psi} \gamma_5 \gamma^\mu D_\mu (\omega) \psi - \overline{D_\mu (\omega)\psi} \gamma_5 \gamma^\mu \psi \right] ,
\end{equation}
and combine it with the gravitational action (\ref{gravaction2}) in the presence of fermions.
One then observes~\cite{mercuri} that variations of the action with respect to the irreducible
components of the contorsion tensor, $T^\mu$, $S^\mu$ and $q^{\mu\nu\rho}$ yield
\begin{equation}\label{tsmerc}
T^\mu = \frac{3}{4} \eta \left(\frac{\gamma/\eta - \gamma^2}{\gamma^2 + 1}\right) J_{(A)}^\mu~, \qquad
S^\mu = - 3 \gamma \, \eta \frac{\gamma/\eta + 1}{\gamma^2 + 1} J_{(A)}^\mu~, \qquad q^{\mu\nu\rho} = 0~.
\end{equation}
The inconsistency with the Lorentz properties of the first equation, involving the trace vector $T^\mu$,
is thereby avoided in the limit
\bea\label{limitetagamma}
\eta \rightarrow \frac{1}{\gamma}~,
\eea
in which case the torsion assumes the form of the Einstein-Cartan theory (\ref{torsionec}).
However, the important point is that the limit (\ref{limitetagamma})
may be taken for {\it any} value of the Immirzi parameter $\gamma$.
In this limiting case, as noted in~\cite{mercuri}, the
modified Holst action is nothing but a total derivative, and can be expressed solely in terms of
\emph{topological invariants}, namely the
so-called Nieh-Yan invariant density~\cite{ny}, and a \emph{total divergence} of the fermion axial current:
\begin{eqnarray}\label{ny}
S_{\rm holst} &=& -i\frac{\eta}{2} \int d^4 x \left[I_{NY} + \partial_\mu J_{(A)}^\mu \right]~, \nonumber \\
I_{NY} & = & \epsilon^{\mu\nu\rho\sigma} \left[ C_{\mu\nu}^ a C_{\rho \sigma a } - \frac{1}{2} \Sigma_{\mu\nu}^{\,\,\ a b}R_{\rho \sigma a b } \right] = \epsilon^{\mu\nu\rho\sigma} \partial_\mu C_{\nu \rho\sigma}~,
\end{eqnarray}
where $\Sigma_{\mu\nu}^{\,\,\ a b} = \frac{1}{2} e^a_{\left[\mu \right.} e^b_{\left. \nu \right]}$ and
$C_{\mu\nu\rho}$ is the contorsion tensor defined previously. Notice that in the last equality for the
Nieh-Yan topological invariant we took into account the fact that, for the torsion (\ref{torsionec}), the
term quadratic in the contorsion in $I_{NY}$ vanishes:
$\epsilon^{\mu\nu\rho\sigma}  C_{\mu\nu}^ a C_{\rho \sigma a }=0$.

This implies that in this limit, variations of the full gravitational action
with respect the connection will not affect the equations of motion.
However, there is an axial- current-current term in the effective action,
which is independent of the Barbero-Immirzi parameter~\cite{mercuri,alexandrov,mercuri2}.
This term is a \emph{repulsive} interaction of the form:
\begin{equation}\label{ec}
S_{\rm int} = - \int d^4 x e \frac{3}{2}\pi G_N {J_{(A)}}_a J^{a}_{(A)}~,
\end{equation}
which coincides with the corresponding term in the Einstein-Cartan theory.
Thus, the case $\gamma = \pm i$ is incorporated trivially, as the effective action
turns out to be independent of the parameter $\gamma$ in this limit. 

The topological nature of the Barbero-Immirzi parameter at the classical level has been clarified in~\cite{sengupta}
via a canonical Hamiltonian analysis. As pointed out in~\cite{seng}, however, subtleties arise at the quantum level. 
Specifically, the standard Dirac quantization procedure for solving  
the second-class constraints that characterise the relevant models before quantization proves insufficient to 
preserve the topological nature of the Barbero-Immirzi parameter. The Nieh-Yan invariant density (\ref{ny}) 
vanishes `strongly' in this case after implementation of the constraints. Nevertheless, alternative procedures for 
quantization have been suggested~\cite{seng}, in which the elimination of the second-class constraints before 
quantization is avoided, and thus the Nieh-Yan density is non-vanishing. This leads~\cite{seng} to a consistent 
topological interpretation of the Barbero-Immirzi parameter, but also indicates the subtle differences
between this parameter and the QCD $\theta$ (instanton) angle.

\subsection{Promotion of the Barbero-Immirzi parameter to an axion field \label{sec:axion}}

The promotion of the Barbero-Immirzi parameter to a space-time dynamical field was proposed in~\cite{bifield},
and a canonical formalism for its quantization in non-supersymmetric theories was developed in~\cite{bifield2}.
In this approach, the total divergence of the Nieh-Yan topological invariant (\ref{ny})
acquires dynamical meaning, resulting in a \emph{pseudoscalar field}
replacing the constant Barbero-Immirzi parameter. The field is pseudoscalar because the
Barbero-Immirzi parameter couples to the dual of the curvature tensor. A canonical kinetic term for the induced
field $\phi$ is obtained if we define it in terms of the Barbero-Immirzi field $\gamma (\vec{x}, t)$ by~\cite{bifield}
\begin{equation}
\phi = \sqrt{3} {\rm sinh}^{-1}(1/\gamma) ,
\end{equation}
which implies that the on-shell gravitational equations in this case become those obtained from the
Einstein-Hilbert action in the presence of a scalar field:
\begin{equation}
\mathcal{G}_{\mu\nu} = \kappa^2 \left( (\partial_\mu \phi) (\partial_\nu \phi) - \frac{1}{2}g_{\mu\nu} (\partial^\rho \phi) (\partial_\rho \phi) \right)~,
\end{equation}
where $\kappa^2=8\pi G_N = 8\pi/M_P^2 $ is the gravitational constant, with $M_P$ the four-dimensional Planck mass,
and $\mathcal{G}_{\mu\nu}$ is the standard Einstein tensor.
The equation of motion of the field $\phi$ is the standard Klein-Gordon equation $\Box \phi = 0$, which
includes the case of a constant Barbero-Immirzi parameter as a trivial solution in which the 
standard general relativity equations are recovered.

It should be noted that consistency of the canonical formulation of this field extension of the Barbero-Immirzi
parameter~\cite{bifield2} requires that in the Ashtekar-Barbero connection only a constant Barbero-Immirzi
parameter enters, which may be identified with a vacuum expectation value of the Barbero-Immirzi pseudoscalar field.
The pseudoscalar nature of the Barbero-Immirzi field makes it resemble an axion. The feature that this field appears
in the effective action only through its derivatives was argued in~\cite{bifield} to be essential for realizing
the Peccei-Quinn U(1) symmetry characteristic of general axion fields.

\subsection{Local supersymmetry (supergravity)  and Barbero-Immirzi terms}

The above Holst framework may be extended to supergravity actions, with the corresponding
supersymmetries being preserved in the limit $\gamma \rightarrow 0$, as in the case of global supersymmetry.
In the context of $N=1$ supergravity~\cite{ferrara}, a generalization to include a Holst term with a
Barbero-Immirzi parameter $\gamma$ would break the underlying local supersymmetry, except in the
limits $\gamma \rightarrow 0$ or $\infty$. In this case, following the standard procedure of varying the Holst action
with respect the spin connection $\omega$ would yield a torsion contribution to the total $\omega$,
involving the gravitino axial current.
However, the supersymmetry of such an action can be preserved by modifying the Holst action
by the addition of appropriate fermion bilinears that are total derivatives, expressible in terms of the
corresponding Nieh-Yan invariant densities, as in the non-supersymmetric case (\ref{ny}) discussed
above~\cite{mercuri}. In this way, Barberi-Immirzi terms do not affect the equations of motion that satisfy the
local $N=1$ supergravity transformations~\cite{ferrara}. 

As discussed in~\cite{tsuda}, one may construct an
$N=1$ supergravity version of the Holst action as follows. One first adds the Holst action to the purely
gravitational (Einstein-Hilbert) sector of the theory:
    \begin{equation}\label{sgholst}
    \mathcal{L}_{\rm G} = \frac{1}{16\pi G_N} e\, e^\mu_a e^\nu _b \left(R^{a b}_{\,\,\,\mu \nu} - \frac{1}{\gamma} \epsilon^{a b}_{\,\,\, c d} R^{c d}_{\,\,\, a b} \right) ~, \quad e \equiv {\rm det}(e^\mu_a)~,
    \end{equation}
    where $R^{a b}_{\,\,\, \mu \nu} = \partial_{\left[ \mu \right.}\omega^{ab}_{\left. \nu \right]} + \omega_{\left[ \mu \right.}^{a c} \omega_{\left. \nu \right]}^{c b} $  is the curvature tensor obtained from the connection $\omega_\mu^{a b} $,
which includes torsion when the system couples to fermions, and $\gamma$ is the Barbero-Immirzi parameter,
which is in general complex.
Next, one elevates this to the $N=1$ supergravity Holst action~\cite{tsuda} by coupling the gravitational action to the
ordinary Lagrangian for a Majorana Rarita-Schwinger (RS) spin-$3/2$ fermion field $\psi_\mu$,
plus a total derivative of the axial gravitino current density, proportional to a (complex) parameter $\eta$.
In flat space-time this recipe would give~\cite{tsuda}
    \begin{eqnarray}\label{rseta}
    \mathcal{L}_{\rm RS} = L_{\rm RS} ({\rm ordinary}) + \frac{i}{4} \partial_\mu \left(\epsilon^{\mu\nu\rho\sigma} \overline{\psi}_\nu \gamma_\rho \psi_\sigma\right) =
    \epsilon^{\mu\nu\rho\sigma} \overline{\psi}_\mu \gamma_5 \gamma_\rho \frac{1 - i \eta \gamma_5}{2}\partial_\sigma \psi_\nu ~.
    \end{eqnarray}
The coupling to gravity is achieved by the usual minimal prescription of replacing ordinary derivatives in flat space by gravitationally-covariant derivatives containing the (torsionful) connection:
\begin{equation}
\partial_\sigma \rightarrow D_\mu \equiv \partial_\mu + \frac{i}{2} \omega_{a b \, \mu}\sigma^{a b}~, \quad \sigma^{a b} \equiv \frac{i}{4} [{\tilde \gamma}^a, \, {\tilde \gamma}^b ]~,
\end{equation}
so that the gravitational RS Lagrangian reads:
\begin{eqnarray}\label{grs}
\mathcal{L}_{\rm GRS} = \epsilon^{\mu\nu\rho\sigma} \overline{\psi}_\mu \gamma_5 \gamma_\rho
\frac{1 - i \eta \gamma_5}{2} \, D_\sigma \psi_\nu ~.
\end{eqnarray}
The $N=1$ supergravity Lagrangian is then obtained by
adding the Holst gravitational action (\ref{sgholst}) to (\ref{grs}):
\begin{eqnarray}\label{etagamma}
\mathcal{L}_{\rm N=1~SG} & = &  \frac{1}{16 \pi G_N} \int d^4 x \left[ e \Sigma^{\mu\nu}_{\,\, a b} R_{\mu\nu}^{\,\,\, ab}(\omega) -
\epsilon^{\mu\nu\rho\sigma} \overline{\psi} \overline{\psi}_\mu \gamma_5 \gamma_\rho \, D_\sigma (\omega) \psi_\nu \right] \nonumber \\
& + & \frac{i}{16\pi G_N} \int d^4 x \left[ \frac{1}{\gamma} \, e \Sigma^{\mu\nu}_{\,\, a b} \tilde{R}_{\mu\nu}^{\,\,\, ab}(\omega) -
\eta \, \epsilon^{\mu\nu\rho\sigma} \overline{\psi} \overline{\psi}_\mu  \gamma_\rho \, D_\sigma (\omega) \psi_\nu \right]~, \nonumber \\
{\rm where} & & \Sigma_{\mu\nu}^{a b} \equiv \frac{1}{2} e^a _{\left[\mu\right.} e^b _{\left.\nu\right]}~, \quad
\tilde{R}_{\mu\nu}^{a b} \equiv \frac{1}{2} \epsilon^{a b c d} \, R_{\mu \nu c d }~.
\end{eqnarray}
Following \cite{tsuda}, one may vary the action with respect to the following quantity constructed out of the
torsionful connection $\omega_\mu^{a b}$:
\begin{equation}
B_{a b \mu} \equiv \frac{1}{2} \left(\omega_{a b \mu } - \frac{1}{2\gamma} \epsilon_{a b}^{\,\,\, c d} \omega_{c d \mu} \right) ,
\label{genconnection}
\end{equation}
obtaining:
\begin{eqnarray}\label{gravitino}
D_\mu \left( e e^\mu_{\left[ a \right.} e^\nu_{\left. b \right]} \right) & = & \frac{\gamma (1 + \eta \gamma)}{\gamma^2 + 1} X_{a b}^\nu + \frac{\gamma \left( 1 - \eta \gamma\right)}{\gamma^2 + 1} \epsilon_{a b }^{\,\,\, cd} X_{cd}^\nu  \nonumber \\
& = & \left(1 - \frac{1 - \eta \gamma}{\gamma^2 + 1} \right) X_{a b}^\nu + \frac{\gamma \left( 1 - \eta \gamma\right)}{\gamma^2 + 1} \epsilon_{a b }^{\,\,\, cd} X_{cd}^\nu~, \nonumber \\  {\rm where} & \quad & X_{a b}^\mu \equiv \frac{1}{1 + \eta^2} \left(\frac{\delta \mathcal{L}_{RS}}{\delta \omega^{a b}_{\,\,\, \mu}} + \frac{\eta}{2} \epsilon_{a b}^{\,\,\, cd} \frac{\delta \mathcal{L}_{RS}}{\delta \omega^{c d}_{\,\,\, \mu}}\right) = \frac{i}{4} \epsilon^{\mu\nu\rho\sigma} \overline{\psi}_\mu \gamma_5 \gamma_\rho \sigma_{ab} \psi_\sigma~.
\end{eqnarray}
The reader will notice that (\ref{gravitino}) has formal analogies with the Dirac fermion case (\ref{defermi}),
in that similar structures appear containing the corresponding fermion current terms. In this case they
are related by Fierz identities to terms containing the gravitino axial current
$J^\mu_{\rm gravitino} = \frac{1}{2}\epsilon^{\mu\nu\rho\sigma} \overline{\psi}_\nu \gamma_\rho \psi_\sigma$.

As in (\ref{defermi}), we observe that, for the purely imaginary Barbero-Immirzi parameters $\eta$ and $\gamma$
of interest to us here, the coefficient of the second term in the right-hand-side of (\ref{gravitino}) is purely imaginary,
and so does not contribute to the Hermitean effective Rarita-Schwinger action, leaving only contributions from the
first term. In fact, apart from the overall coefficient, the structure of these four-fermion terms is the same as the
torsion-induced four-fermion terms in standard $N=1$ supergravity~\cite{ferrara}.
The usual four-fermion terms of $N=1$ supergravity in the second-order formalism are obtained in the limit
\begin{equation}\label{sugra}
\frac{\eta}{\gamma} -1 \rightarrow 0~,
\end{equation}
in which case the Lagrangian (\ref{etagamma}) differs from the standard $N=1$ supergravity Lagrangian simply
by the total derivative of the axial gravitino current:
\begin{equation}
\mathcal{L} = \mathcal{L}_{N=1} ({\rm second~order~formalism}) + \frac{1}{4\gamma} \partial_\mu \left(\epsilon^{\mu\nu\rho\sigma} \overline{\psi}_\nu \gamma_\rho \psi_\sigma \right)~.
\end{equation}
In the limit $\gamma \eta = 1$, the Holst modification of the gravity action, with the fermionic corrections,
is nothing but the topological invariant Nieh-Yan density plus the total derivative of the gravitino current,
in analogy with the Dirac fermion case (\ref{ny}) discussed previously~\cite{kaul}:
\bea
\label{nysugra}
S_{N=1~{\rm Holst}} = - \frac{1}{2 \gamma} \int d^4 \left[I_{NY} + \frac{1}{2} \partial_\mu \left(\epsilon^{\mu\nu\rho\sigma} \overline{\psi}_\nu \gamma_\rho \psi_\sigma\right)\right]~,
\eea
with $I_{NY}$ given in terms of the torsion tensor as in (\ref{ny}).

In the case of N=1 supergravity, the torsion $C_\mu^{a b}$, given as a solution of (\ref{gravitino}) for $\eta \gamma = 1$, leads to the following contorsion tensor~\cite{ferrara}:
\begin{equation}\label{torsionsugra}
C_{\mu\alpha\beta} (\psi) = \frac{1}{4} 8 \pi G_N \left( \overline{\psi}_\alpha \gamma_\mu \psi_\beta
+ \overline{\psi}_\mu \gamma_\alpha \psi_\beta - \overline{\psi}_\mu \gamma_\beta \psi_\alpha \right) ,
\end{equation}
with the torsion given by $T^\lambda_{\mu \nu} = -\frac{1}{2} C_{\left[\mu \nu \right]}^{\,\,\,\lambda}$, such that:
\begin{equation}
D_{\left[ \mu \right.}(\omega)e^a_{\left. \nu \right]} = 2 T_{\mu\nu}^a = \frac{1}{2}\overline{\psi} \gamma^a \psi_\nu .
\end{equation}
With such a torsion, an appropriate Fierz rearrangement implies
$\epsilon^{\mu\nu\alpha\beta} T_{\mu\nu}^a T_{\alpha\beta \, a} =0$
and the Nieh-Yan invariant can be expressed as a total derivative, exactly as in the Dirac spin-1/2 fermion case (\ref{ny}).
As such, the $N=1$ supergravity equations of motion are not affected, and the on-shell local supersymmetry
transformations remain intact for arbitrary values of the parameter $\gamma$.
Explicitly, the action of Barbero-Immirzi-modified N=1 supergravity in the limit $\gamma= 1/\eta$,
with $\gamma $ \emph{arbitrary}, is invariant under the following transformations generated by a Majorana local
parameter $\alpha (x)$~\cite{tsuda}:
\bea\label{nsugraimmtrns}
\delta \psi_\mu &=& \frac{1}{\kappa} D_\mu \alpha ~, \nonumber \\
\delta e^a _\mu & = & \frac{i \kappa }{2} \overline{\alpha} \gamma^a \psi_\mu~, \nonumber \\
\delta B_{a b \mu} &=& \frac{1}{2} \left(\mathcal{C}_{\mu a b} - e_{\mu \left[ a \right.} \mathcal{C}^c _{\left. c b\right]}\right)~,  \nonumber \\ {\rm with} \quad \mathcal{C}^{\mu\nu\rho} &\equiv & e^{-1} \kappa \epsilon^{\nu\rho\alpha\sigma}\overline{\alpha}\gamma_5 \gamma^\mu \frac{1 - \frac{i}{\gamma}\gamma_5 }{2} D_\alpha \psi_\sigma~,
\eea
where $\kappa^2 = 8\pi G_N$ is the gravitational coupling. The limit $\gamma \rightarrow \pm i$
(purely imaginary Immirzi parameter)
leads to chiral $N=1$ supergravity. Substitution of the torsion (\ref{torsionsugra}) into the first-order action yields
the standard $N=1$ supergravity four-fermion interaction terms~\cite{ferrara}. The existence of torsion-induced
four-gravitino terms in the effective Lagrangian in the $\eta \gamma = 1$ supergravity limit, for \emph{arbitrary values}
of the Immirzi parameter $\gamma$, is analogous to the axial current-current interactions in the
Einstein-Cartan theory (\ref{intec})~\footnote{Extensions of this formalism to $N=2$ and $N=4$ supergravity models are also known~\cite{kaul},
but we do not discuss them in the current article.}$^{,}$\footnote{The interpretation of the Barbero-Immirzi parameter 
as a topological parameter in general supergravity theories has been confirmed by a detailed canonical analysis of the 
spin-$3/2$ fermionic action in~\cite{kaulseng}.}.

We next remark that the promotion of the Barbero-Immirzi parameter to an axion field,
as discussed in subsection \ref{sec:axion} above,
can also be applied to $N=1$ four-dimensional supergravity, with the axion field being promoted to a complex chiral
axion-dilaton superfield, whose lowest component comprises a scalar and a pseudoscalar field,
which play the r\^oles of the dilaton and axion respectively~ \cite{gates}. The incorporation of a Barbero-Immirzi
field in such a formalism can be done neatly by first \emph{complexifying} the gravitational coupling constant
$\kappa^2$ of the standard $N=1$ supegravity theory:
\begin{equation}\label{complexf}
 \frac{1}{\kappa^2} \rightarrow \frac{1}{\kappa^2} \left(1 + i \eta \right) ,
 \end{equation}
where $\eta$ is a real dimensionless parameter identified with the inverse of the Barbero-Immirzi parameter:
$\eta = 1/\gamma$ as discussed above.

As is standard in supersymmetry, such complex couplings are consistent, given that in the pertinent actions
one always includes the complex conjugate, so the final action is real.
The gauge-invariant action of $N=1$ supergravity in superfield formalism reads~\cite{gatesbook}:
\begin{equation}\label{sugrasf}
S_{SG} = -\frac{3}{\kappa^2} \int d^4 x d^2 \theta d^2 {\overline \theta} E^{-1} = -\frac{3}{2\kappa^2}  \int d^4 x d^2 \theta  \mathcal{E}\,\mathcal{R} + h.c.~,
\end{equation}
where in the second equality we used the chiral-superspace formalism of $N=1$ supergravity,
which is convenient for our discussion below~\cite{gates}. In the above formulae, $E^{-1}= {\rm SDet}E^M_A$
is  a supervierbein density in full curved superspace $(x, theta, {\overline \theta})$, with SDet denoting a
superdeterminant~\cite{gatesbook}, while
$\mathcal{E}$ and $\mathcal{R}$ denote supersymmetric generalizations of the volume element and the
Lagrangian density in chiral superspace. Details of the formalism and the equivalence of the action (\ref{sugrasf})
with the standard $N=1$ supergravity action~\cite{ferrara} in component formalism are provided
in~\cite{gates,gatesbook} and will not be repeated here.

For our purposes we note that the complexification (\ref{complexf}),
when substituted into the chiral superspace action (\ref{sugrasf}),
yields the supersymmetric Holst term of~\cite{tsuda} (\ref{nysugra}),
which for constant $\eta$ is a total derivative, thereby not affecting the $N=1$ supergravity equations of motion.
Promotion of the parameter $\eta$ to a field is achieved by replacing $\eta$ in
(\ref{complexf}) by a complex chiral superfield $\mathcal{Z}(x, \theta)$ with scalar component
$\mathcal{Z}(x, \theta)| = \varphi (x) + i b(x)$, where $\varphi (x)$ is the dilaton scalar and
$b(x)$ is a four-dimensional axion pseudoscalar field~\cite{gates}:
\begin{equation}
-\frac{3}{2\kappa^2}\left(1 + i \eta \right) \rightarrow \mathcal{Z}(x, \theta) ,
\end{equation}
so that the $N=1$ supergravity action becomes:
\begin{equation}\label{sugracompl}
\int d^4 x d^2 \theta  \mathcal{E}\, \mathcal{Z}\,\mathcal{R} + h.c.~.
\end{equation}
Thus, in this formalism the field $\mathcal{Z}$ plays the r\^ole of a `Lagrange multiplier' superfield,
whose variations
yield the dilaton-axion field equations of motion. These include the constant
Barbero-Immirzi case ($\eta $ = const) as
a trivial solution, in analogy with the non-supersymmetric example discussed above.

Following the argument of  \cite{gates}, we note next that, under superfield Weyl transformations
(which comprise ordinary Weyl transformations of component fields, chiral rotation and a superconformal
symmetry transformation), the following transformation laws are obeyed by the superfields entering (\ref{sugracompl}):
\begin{equation}
\mathcal{E} \rightarrow e^{3\Phi} \mathcal{E}~, \quad \mathcal{R} \rightarrow e^{-2\Phi}\left(\mathcal{R} - \frac{1}{2}{\overline \nabla}^2\right)e^{\overline{\Phi}} ,
\end{equation}
where $\Phi$ is an arbitrary covariantly-chiral superfield $\overline{\nabla}_{\cdot \alpha} \Phi = 0$,
with $\nabla_\alpha$ denoting a curved superspace covariant derivative.
The above freedom under Weyl transforms allows the imposition of a holomorphic
`gauge fixing':
\begin{equation}\label{gauge}
\mathcal{Z} = \Phi
\end{equation}
as the simplest condition. Other arbitrary functions of $\Phi$ are allowed in more complicated gauge fixings.
Such ambiguities may be fixed dynamically when one considers the embedding of the
effective action (\ref{sugracompl}) in a more microscopic framework such as string theory.

The use of (\ref{gauge}) results in the following form of the supergravity action,
involving a chiral complex superfield coupled to supergravity
(we revert here to the full curved superspace notation $(x, \theta, \overline{\theta})$):
\begin{equation}\label{imsugra}
S_\Phi = \int d^4 x d^\theta d^2{\overline \theta} E^{-1} e^{\Phi + \Phi} \left(\Phi + \overline{\Phi}\right) \equiv -3\int d^4 x d^\theta d^2{\overline \theta} e^{-K/3} ,
\end{equation}
with a K\"ahler potential
\begin{equation}\label{kahler}
K(\Phi, \overline{\Phi}) = -3{\rm ln}\left(-\frac{1}{3}\,e^{\Phi + \Phi}\, [\Phi + \overline{\Phi}]\right)~.
\end{equation}
In general such a simplified form may be modified by more complicated potentials of the dilaton-axion superfields,
when such models are viewed as low-energy approximations to some string theory. In particular, when
higher-order string loop corrections are taken into account, non-trivial corrections to the dilaton/axion
potentials may be generated, that stabilize the fields to constant values.

\subsection{Global supersymmetry breaking and the Immirzi dilaton/axion superfield}

We now discuss how local supersymmetry may be broken by the superfield $\Phi$
acquiring an appropriate vacuum expectation value. We first describe some generic considerations on
supersymmetry breaking due to chiral matter superfields.
In general there are two generic types, $F$ and Fayet-Iliopoulos $D$-term breaking,
the former arising when the $F$-term of a chiral superfield acquires a non-trivial vacuum
expectation value $<F> = f \ne 0$, and we start with this case.
Our initial considerations below refer to generic chiral superfields, and we turn later to the
specific case of the dilaton/axion superfield.

The $F$-term breaking of supersymmetry implies in general~\cite{komar} that the
Ferrara-Zumino (FZ) current superfield multiplet
$\mathcal{J}^{\alpha {\dot \alpha}} \equiv \mathcal{J}_\mu \sigma^{\mu \, \alpha {\dot \alpha}}$
is well defined in such a scenario (we use standard superfield notation in what follows).
This (Lorentz vector) multiplet consists of the supersymmetry current $S_{\mu \alpha}$,
the energy-momentum tensor $T_{\mu\nu} = T_{\nu\mu}$,
and an $R$-symmetry current, which is not necessarily conserved.
The current obeys a conservation equation
\begin{equation}\label{current}
{\overline D}^{\dot \alpha} \mathcal{J}_{\alpha {\dot \alpha}} = D_\alpha \Phi ,
\end{equation}
where $\Phi$ is a complex chiral superfield, which will in our case be the dilaton-axion multiplet~%
\footnote{We note that the dilaton is a real field, whereas we use here complex supermultiplets,
which have complex scalar fields with two degrees of freedom in the lowest component of the
superfield. As we discuss later in the article,
we identify one of these with the usual dilaton while the other, which
couples in the supergravity sector with the dual of the curvature tensor,
plays the role of a pseudoscalar axion.}.
In the absence of $D$-term supersymmetry breaking, the FZ current multiplet is gauge invariant.

The existence of such a well-defined FZ current multiplet implies, by means of generic supersymmetry
considerations~\cite{komar}, that
at low energies the solution of the system of supersymmetry transformations of the infrared limit of the
chiral superfield $\Phi_{NL} = \phi_{NL} + \theta^\alpha \Psi_{NL\, \alpha} + \theta^\alpha \theta_\alpha F_{NL}$
leads to the following expression:
\begin{equation}  \label{solution}
\Phi_{NL} = \frac{G^2}{2F} + \sqrt{2} \theta G + \theta^2 F ,
\end{equation}
which leads to the constraint
\begin{equation}\label{constr}
\Phi_{NL}^2 = 0 .
\end{equation}
A general discussion in~\cite{komar} established that the fermionic component of the chiral superfield
plays the r\^ole of the Goldstino, and that at low energies its action is the Volkov-Akulov action of
non-linear supersymmetry~\cite{va}.
Thus, in $F$-type supersymmetry breaking the Goldstino always resides in a chiral superfield.

In the Immirzi extension of $N=1$ supergravity (\ref{imsugra}) with the (simplest) K\"ahler potential (\ref{kahler}),
one may assume that the $F$-term of the dilaton/axion Immirzi superfield breaks global supersymmetry,
in which case the low-energy limit will lead to the non-linear constraint (\ref{constr}).
Taking this constraint into account implies that the action (\ref{imsugra}),
in the \emph{flat space-time} limit we start with, assumes the form:
\begin{equation}\label{flat}
S_\Phi = \int d^4 x \mathcal{L_{\rm flat\, \Phi}}~, \quad \mathcal{L_{\rm flat\, \Phi}} = 2\,\left(\int d^2 \theta d^2{\overline \theta}
\Phi \, {\overline \Phi} + \int d^2 \theta \frac{1}{2} \Phi + d^2{\overline \theta} \frac{1}{2} \Phi \right)~.
\end{equation}
This has the same form as the (lowest-order) non-derivative infra-red effective action of~\cite{komar}, which resembles
the trivial (free superfield) case of supersymmetry breaking, except that the superfield $\Phi$  is constrained.
Comparing our normalization with that of~\cite{komar}, we see that the supersymmetry-breaking parameter
$f=\frac{1}{2}$ in this case, in units of the gravitational scale $\kappa = \frac{\sqrt{8\pi}}{M_P}=1$. 
In the case of the simplest K\"ahler potential
(\ref{kahler}), it therefore seems that the supersymmetry parameter $f$ is fixed at the gravitational scale (up to a numerical
factor $1/2$), which is to be expected, given that the Barbero-Immirzi field is a gravitational effect.
However, this might be problematic from the point of view of low-energy (infrared) phenomenology, 
where one might like the supersymmetry breaking scale to be far below the Planck mass.
However, it goes without saying that the embedding of the Barbero-Immirzi model in string theory,
which would entail a more complicated potential and higher-order curvature terms, could change this condition,
as we discuss later on.

In our particular Barbero-Immirzi model, it is the fermionic partner of the dilaton-axion (the `dilatino/axino') that can be
identified with the Goldstino field at low energies, where the constraint (\ref{constr}) is satisfied,
Following the generic supersymmetry analysis of~\cite{komar},
i.e., substituting the solution (\ref{solution}) with $f=1/2$ into (\ref{flat}), the effective Lagrangian for the
Golsdstino field $G_\alpha$ has the form of the Volkov-Akulov
effective Lagrangian for the non-linear realization of supersymmetry~\cite{va}:
\begin{equation}\label{va2}
\mathcal{L}_{\rm IR \, \Phi} = -\frac{1}{4} + i \partial_\mu {\overline G} \overline{\sigma}^\mu G + \overline{G}^2 \partial^2 G^2 - 4 G^2 {\overline G}^2 \partial^2 G^2 \partial^2 {\overline G}^2~.
\end{equation}
In the four-component formalism, this is equivalent to the original Volkov-Akulov
Lagrangian~%
\footnote{In the original works of~\cite{va},
the Lagrangian is written in terms of a four-component Majorana spin-$1/2$ Goldstino field (called $\lambda $):
$\mathcal{L}_\lambda = -(f^2){\rm det}\left(\delta^\mu_\nu + i \frac{1}{2f^2}\overline{\lambda}  \gamma^\nu \partial_\mu \lambda \right)$ that, upon expansion of the determinant and fermionic truncation, yields (\ref{va2}) when one passes into the
two-component spinor formalism. The constant $f$ expresses the strength of global supersymmetry breaking and,
as mentioned above, in our case this occurs at $f=1/2$ in Planck units. The Lagrangian is characterized by a non-linear realization of global supersymmetry with infinitesimal parameter $\alpha$:
$\delta \lambda = f\,\alpha + i \frac{1}{f} \overline{\alpha} \gamma^\mu \lambda \partial_\mu \lambda~.$}.

The above low-energy considerations can be extended~\cite{komar2} to the case of
$D$-term supersymmetry breaking in a rather non-trivial way. In such a case, the FZ multiplet is not well-defined,
in the sense that it is not gauge invariant. This is not a pathology of the theory, it just means that the above
considerations based on the definition of a FZ current cannot apply. Nevertheless, in the low-energy limit one
can still define constrained dilaton superfields $\Phi_{NL}$, so that the connection of the dilatino-axino with the
Goldstino can still be maintained.
In fact, in the string literature~\cite{arkani} there are dynamical supersymmetry models with anomalous
U(1) symmetry and anomaly cancellation by the Green-Schwarz mechanism, where the $D$ terms may be
significant, depending on how the dilaton is stabilized.
In such scenarios, the dilatino also emerges as (mostly) the Goldstino, as in the pure $F$-term case discussed above.

One can discuss generalizations of the FZ current mutliplet to another supermultiplet, also containing the
stress-energy tensor and the supersymmetry current in its components, whose conservation equation involves
necessarily massless chiral scalar superfields. In string theories the r\^ole of such a massless superfield is played
by the dilaton and other moduli fields~\cite{komar2}.
From our point of view, therefore, the incorporation of dilaton-axion superfields in a Jordan-frame-like
modification (\ref{imsugra}) of the standard supergravity action fits into the above picture. In the low-energy
regime of more general supergravity theories, where a FZ multiplet cannot be defined,
condensates of the Goldstino field (which is the dilatino) still form in the infrared, and correspond to the infrared
dilaton (and axion) fields.
It is not clear, however, that such theories are consistent theories of quantum gravity,
given that
they are characterized by additional global $R$ symmetries. Thus, for our purposes below we concentrate on
theories with well-defined FZ multiplets, and therefore primarily $F$-type supersymmetry-breaking models.

\subsection{Coupling to supergravity}

The coupling of the Goldstino to supergravity generates a mass for the gravitino through the absorption of
the Goldstino, via the super-Higgs effect envisaged in~\cite{deser}.
According to this model, the $N=1$ supergravity theory is coupled to a Volkov-Akulov
Majorana fermion that may arise from some spontaneous or dynamical breaking of supersymmetry.
In our extension of $N=1$ supergravity to include a Barbero-Immirzi field,
the coupling is provided by the Lagrangian (\ref{imsugra}) which, in view of the above discussion on the
identification of the dilatino as the Goldstino field in the case of broken supersymmetry, has
at low energies the component form suggested in the super-Higgs effect scenario of~\cite{deser},
namely a non-linear Volkov-Akulov Lagrangian coupled to $N=1$ supergravity.
It is instructive, and illuminating for what follows, to review explicitly this coupled system in component form~%
\footnote{We repeat that in superfield language this is just (\ref{imsugra},
upon using the condition $\Phi^2_{NL} =0$ (\ref{constr})).}.

Thus, we consider a spontaneously-broken supersymmetric theory with a Majorana Goldstino $\lambda$,
whose action takes the non-linear form considered by Volkov and Akulov~\cite{va,deser}:
\bea\label{goldstino}
\mathcal{L}_\lambda = -f^2 {\rm det}\left(\delta^\mu_\nu + i\, \frac{1}{2 f^2} \, \overline{\lambda} \gamma^\mu \partial_\mu \lambda \right) = - f^2 - \frac{1}{2} i \overline{\lambda} \gamma^\mu \partial_\mu \lambda + \dots
\eea
Here we keep the discussion general by allowing for an arbitrary value of the
parameter $f$, as is possible in general models: see the discussion above.
As discussed in~\cite{deser}, one can promote the global supersymmetry to a local one,
by allowing the parameter $\alpha (x)$ to depend on space-time coordinates,
and coupling the action (\ref{goldstino}) to that of $N=1$ supergravity in such a way that
the combined action is invariant under the following supergravity transformations:
\bea\label{sugratransgolds}
\delta \lambda &=& \beta^{-1} \alpha (x) + \dots ~, \nonumber \\
\delta e^a _\mu  & = & -i \kappa \overline{\alpha}(x) \gamma^a \psi_\mu~, \nonumber \\
\delta \psi_\mu & = & - 2 \kappa^{-1} \partial_\mu \alpha (x) + \dots
\eea
The action that changes by a divergence under these transformations is the standard $N=1$ supergravity action plus
\begin{equation}\label{va2b}
L_\lambda = - f^2 e - \frac{i}{2}\overline{\lambda} \gamma^\mu \partial_\mu \lambda - \frac{i\,f}{\sqrt{2}} \overline{\lambda} \gamma^\nu \psi_\nu + \dots ,
\end{equation}
which contains the coupling of the Goldstino to the gravitino.
The Goldstino can be gauged away~\cite{deser} by a suitable redefinition of the gravitino field and the tetrad.
One may impose the gauge condition
\begin{equation}\label{gravinogauge}
\psi_\mu \gamma^\mu = 0 ,
\end{equation}
but this leaves behind a negative cosmological constant term, so the total Lagrangian after these redefinitions reads:
\begin{equation}\label{va3}
\mathcal{L}_{\rm eff} = -f^2e + (N=1~{\rm supergravity}) .
\end{equation}
The presence of four-gravitino interactions in the standard $N=1$ supergravity Lagrangian
in  the second-order formalism, due to the fermionic contributions to the torsion in the spin connection
(\ref{torsionsugra}), implies also an induced  gravitino mass term that is generated dynamically.

To see this, one may simply linearize the appropriate four-gravitino mass terms of the $N=1$ supergravity
Lagrangian in the second-order formalism~\cite{ferrara},
by means of an auxiliary scalar field $\rho (x)$~\cite{smith}:
\begin{eqnarray}\label{effactionlinear}
\mathcal{L}{\rm eff} =
-\frac{1}{\kappa^2} R(e) + \frac{1}{2}\epsilon^{\mu\nu\rho\sigma} \overline{\psi}_\mu \gamma_5 \gamma_\nu D_\rho \psi_\sigma + \rho^2(x) - \sqrt{11} \kappa \rho (x) \left(\overline{\psi}_\mu \Gamma^{\mu\nu} \psi_\nu \right) + \dots
\end{eqnarray}
with $\Gamma^{\mu\nu} \equiv \frac{1}{4} [ \gamma^\mu, \gamma^\nu ]$, where the $\dots $ indicate terms we are not
interested in, including other four-gravitino interactions with $\gamma_5$ insertions, as well as other
standard $N=1$ interactions and auxiliary supergravity fields.
On account of the gauge fixing condition (\ref{gravinogauge}), we have
\begin{equation}\label{identity} 
 \overline{\psi}_\mu \Gamma^{\mu\nu} \psi_\nu = -\frac{1}{2}\overline{\psi}_\mu \psi^\mu ,
\end{equation}
using the anti-commutation properties of the Dirac matrices $\gamma^\mu$. 
The formation of a condensate~\footnote{Sometimes~\cite{deser} the gravitino mass term is defined in the presence of $\gamma_5$ as:
$m \frac{1}{2} \epsilon^{\mu\nu\rho\sigma} \overline{\psi}_\mu \Gamma_{\nu \rho} \psi_\sigma $. It is easy to
adapt to this case by linearizing the appropriate gravitino four-fermion terms that contain the square of such terms,
which we did not indicate explicitly in the effective action (\ref{effactionlinear}). The analysis is exactly the same,
whichever form of gravitino mass we seek to create dynamically.}.
\bea \label{gravinocond}
\langle \rho (x) \rangle \equiv \rho \sim ~\langle \overline{\psi}_\mu \Gamma^{\mu\nu} \psi_\nu \rangle ~,
 \eea
 which should be independent of $x$ because of the translation invariance of the vacuum,
 is possible by minimizing the effective action  (\ref{effactionlinear}) along the lines in~\cite{smith}.

We observe~\cite{deser,smith} that the formation of the condensate may cancel the negative cosmological
constant term. The condensate contributes to the vacuum energy a term of the form
\begin{equation}\label{verho}
\int d^4 x \,e \,\rho^2  > 0~,
\end{equation}
and at tree level one can fine-tune this term so as to cancel the negative cosmological constant
of the Volkov-Akulov Lagrangian (\ref{va2b},\ref{va3}), which depends on the supersymmetry breaking scale $f^2$, by setting:
\begin{equation}\label{condition}
\rho^2 =  f^2~.
\end{equation}
In ref.~\cite{smith2}, a one-loop effective potential analysis has demonstrated that such a cancellation occurs for a suitable value of the parameter $f$.
Whether the situation persists to higher orders, so that the cancellation of the effective cosmological constant can be achieved exactly,
is not known.
Assuming this to be the case, or restricting ourselves to one-loop order, we may therefore consider the quantization of the
gravitino field in a Minkowski space-time background and discuss its dynamical mass generation in the same spirit as in 
the flat space-time prototype case of the chiral-symmetry breaking four-fermion Nambu-Jona-Lasinio 
model~\cite{nambu}~\footnote{It should be remarked here, especially in connection with the D-foam 
model~\cite{Dfoam} discussed in the next Section, that such a cancellation  of the cosmological constant 
may not characterise the microscopic model, which may thus exhibit an (anti-)de Sitter background, 
depending on the net sign of the vacuum energy contributions.
It is well-known~\cite{candelas} that  the dynamical formation of chiral condensates via four-fermion interactions, 
as of of interest to us here, exhibits
better UV behaviour than in the corresponding flat space-time case~\cite{nambu}, in the sense that
any potential UV  infinities, which are regularised by means of an UV  cut-off $\Lambda$, may be absorbed into 
the cosmological constant and Planck scale of the (anti-)de Sitter space-time. For chiral symmetry breaking in the early 
Universe , for instance,  this has been demonstrated explicitly in~\cite{neubert}. In our case, the regularization of UV 
infinities by a cut-off is consistent with the breaking of supersymmetry, and one may attempt a similar analysis as in the 
spin-1/2 four-fermi models of~\cite{neubert}. However, we do not do this in the present work}$^{,}$~\footnote{We also note that
in (anti)de Sitter space-times even repulsive interactions
lead to condensates.}.

Enforcing this cancellation
of the vacuum energy contributions, one may write the effective action for the condensate fluctuations 
$\rho'(x)$, $\rho(x) = \rho + \rho'(x)$ in a standard fashion, by direct substitution in (\ref{effactionlinear}) 
and expansion around a Minkowski space-time, using the vierbein expansion
$e^a_{\,\,\mu} = \delta^a_{\,\,\mu} +  h^a_{\,\,\mu} $ and
taking into account the gauge condition (\ref{gravinogauge}). From the resulting effective action at zeroth order in the 
gravitational field $h$, then, one can read off directly the dynamically-generated gravitino mass as~\cite{smith}:
\begin{equation}\label{ginomass}
m^2_{\rm 3/2} \sim 44 \rho^2  \kappa^2~,
\end{equation}
with $\rho^2 = f^2$, cf (\ref{condition}). Since the graviton remains massless
in this approach, supersymmetry is broken locally.

The size of the condensate $\rho$ can be determined in principle by solving the appropriate gap
equations in the Minkowski  background, following from the expression for the gravitino propagator and the imposition of the vanishing of the tadpole of the fluctuation of the linearising auxiliary field $\rho'(x)$. The analysis of~\cite{smith} leads to
the following gap equation for the gravitino field of $N=1$ supergravity, by passing to the appropriate Fourier $k$-space
in the Minkowski space-time background:
\begin{equation}\label{integral}
\frac{3}{44}\frac{1}{\kappa^2} = \int d^4 k \left(4 + \frac{k^2}{m^2_{3/2}} \right)\frac{1}{k^2 + m^2_{3/2}} ,
\end{equation}
with $m_{3/2}$ given by (\ref{ginomass}). This equation has to be regularised in the UV by a cut-off $\Lambda$, which is consistent with broken supersymmetry. The spherical symmetry of the integrand in (\ref{integral}) yields the following analytic expression for the 
integral~\cite{smith}, assuming a Euclidean formulation and
performing the usual analytic continuation back to Minkowski space-time only at the very end of the computations:
\begin{equation}\label{gap}
\frac{3}{44}\frac{1}{\kappa^2} = \pi^2 \left[ \frac{\Lambda^4}{2m^2_{3/2}} + 3 \Lambda^2 - 3 m^2_{3/2} {\rm ln} \left(\frac{\Lambda^2}{m^2_{3/2}} + 1\right)\right]~.
\end{equation}
The cut-off $\Lambda$ cannot be determined in the low-energy effective field theory framework. In our case, one may use as the UV cut-off $\Lambda$ the supersymmetry breaking scale $f$. However, in our Barbero-Immirzi case as discussed so far, $f$ is of the order of the Planck mass squared, $f= \kappa^2/2 = M_P^2/16\pi$, cf, the discussion following (\ref{flat}) and re-instating units of $M_P$, implying
a gravitino mass of the order of the Planck mass,
as was the case in~\cite{smith}.
Similar results are obtained if one considers the one-loop effective potential analysis 
of~\cite{smith2}, which shows that up to that order
the effective potential for the $\sigma$ field is always positive and vanishes at a non-trivial 
minimum, for a value of the cut-off $\Lambda = \kappa^{-1}$ and values of $f$ and the gravitino mass of similar
(Planckian) order to that indicated by the tree-level analysis described above.

However, these results are not consistent with our
infrared analysis, where we expected a gravitino that is light compared to the Planck mass.
Moreover, there is another problem with such high-scale supersymmetry breaking. For a
dilaton v.e.v.  of the order of the Planck scale, one cannot ignore the quantum fluctuations of the gravitational field
around the classical
anti-de-Sitter metric background,  $g^{0}_{\mu\nu}$, in which the Volkov-Akulov Lagrangian is formulated.
It is for this reason that
the above considerations have been disputed in~\cite{odintsov}.
Indeed, in a linearised gravity approximation, $g_{\mu\nu} = g^{0}_{\mu\nu} + h_{\mu\nu}$,
integrating out the metric fluctuations $h_{\mu\nu}$ in the way suggested in \cite{fradkin} 
leads~\cite{odintsov} to gauge-dependent imaginary parts in the effective action,
indicating an instability of the gravitino condensate~\footnote{However, in our opinion, the gauge-invariant
Batalin-Vilkovisky formalism for the effective action  of $N \ge 1$ supergravity theories, 
as used in \cite{fradkin,odintsov}, is not completely understood at present. Hence,
the gauge dependence of the imaginary parts, and the fact that in some gauges the imaginary parts are 
zero~\cite{odintsov2}, might simply indicate our inadequate understanding of low-energy supergravity theories, 
rather than proving that dynamical mass generation
of a gravitino mass of the order of the Planck mass via four-gravitino interactions is not possible in the above-described fashion. Nevertheless, for our purposes here we are primarily interested in relatively light gravitinos, as already mentioned.}.

Hence it is desirable to be able to extend the Barbero-Immirzi formalism to incorporate light gravitinos, 
corresponding to supersymmetry breaking scales that are low compared to Planck mass,
in which the gravitational fluctuations can be ignored. This could in principle be achieved by embedding
the Barbero-Immirzi effective action (\ref{imsugra}) into a full string theory framework, with an
appropriate dilaton potential to be generated by string loops.
One concrete such framework, that of D-particle space-time foam,  is discussed in the next section,
but other examples may well be possible.  In such a case, a low scale of supersymmetry breaking
$f$ may be introduced, and the problem of fluctuations in the gravitational field could be evaded.
The simplest way of achieving this is to rescale the dilaton/axion superfield field $\Phi$: $\Phi \rightarrow f \Phi$ 
in (\ref{imsugra}, \ref{flat}) and embed the theory into higher dimensions, by assuming for instance a brane 
Universe propagating into a bulk space.
In such a case, the action (\ref{imsugra}) is nothing but part of an effective action on the brane world in the Jordan frame. The 
only ingredient of the Barbero-Immirzi field that we maintain is that it is associated with the dilaton/axion excitation of this 
string theory. There is a potential for the dilaton, field as already mentioned, which determines the global supersymmetry 
properties of the dilaton $F$-term, implying a low $f$.

In such a scenario, the low-energy cut-off $\Lambda$, taken to be of order of the supersymmetry breaking scale $\sqrt{f}$, 
is assumed to be much lower than the four-dimensional Planck scale $M_P$, and also much higher than the 
mass of the gravitino $m_{3/2} \ll \Lambda \sim f \ll M_P$. We then obtain from (\ref{gap}):
 \begin{equation}\label{latestginomass}
m^2_{3/2} \sim \frac{22}{3}\pi^2 \frac{\Lambda ^4}{M_P^2} ~.
\end{equation}
Consistency between (\ref{ginomass}) and (\ref{latestginomass}) is achieved for $\pi^2 \Lambda ^4/3 \sim f^2 $.

We now proceed to discuss a concrete example, that of a D-brane model for
stringy space-time foam, where such a
low supersymmetry breaking scale may be realised.

\section{Quantum Gravity Foam, Dilatons and Supersymmetry Breaking}

\subsection{D-particle foam as a quantum-gravity medium}\label{sec:dfoam}

\begin{figure}[ht]
\centering
\includegraphics[width=9cm]{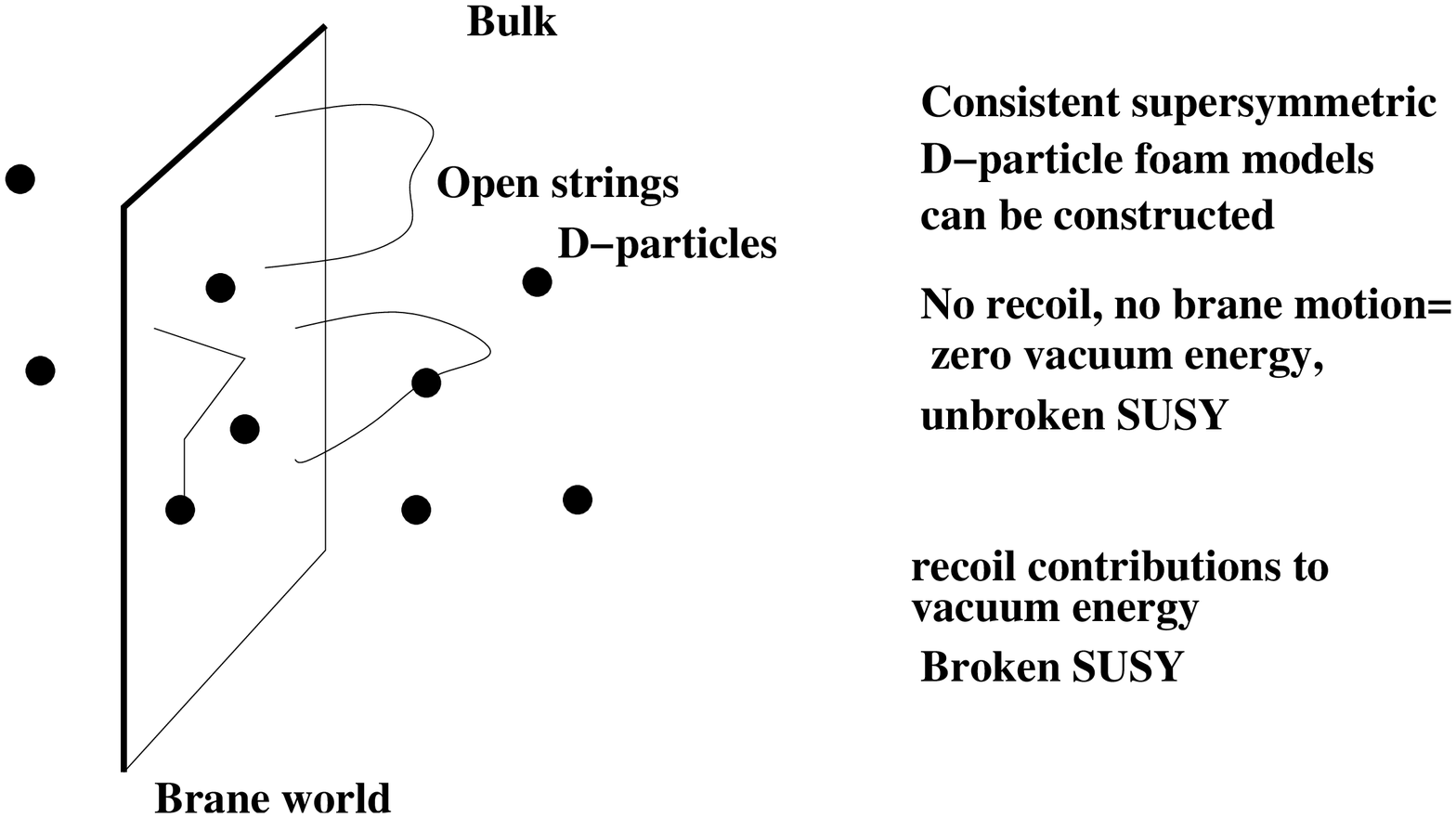} \vfill \vspace{2cm}
\includegraphics[width=9cm]{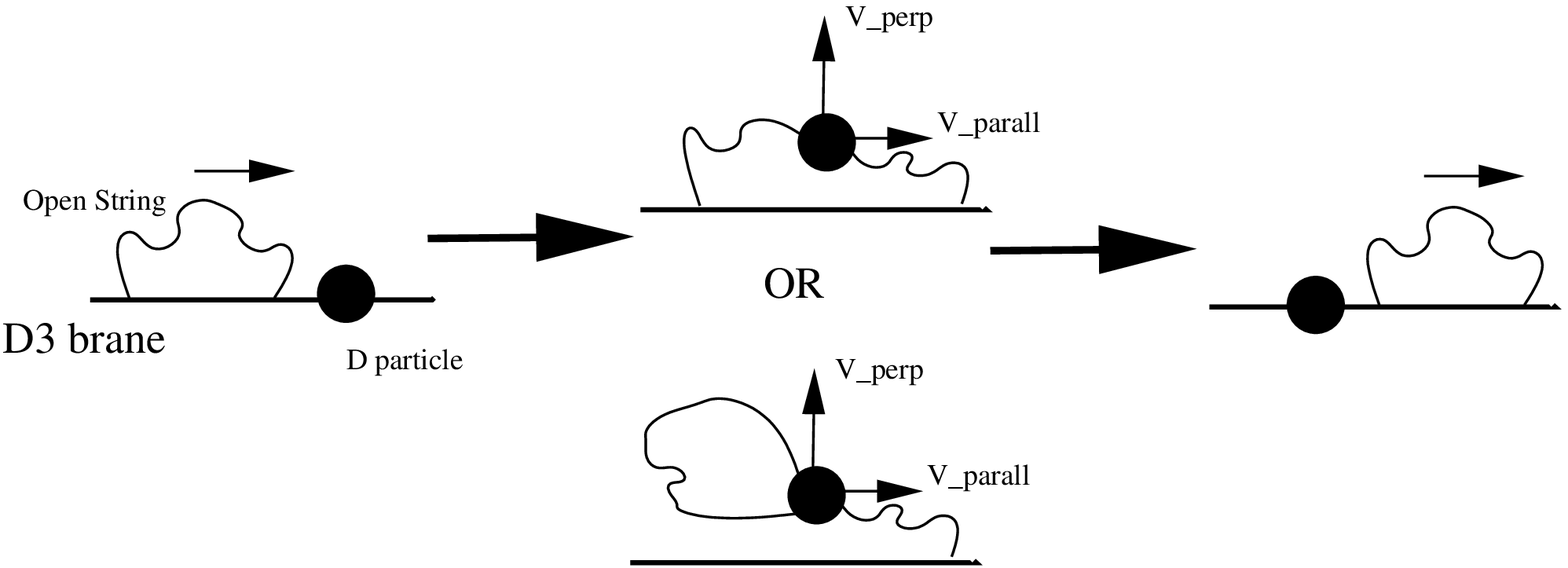}
\caption{\it \underline{Upper}: Schematic
representation of a generic D-particle space-time foam model, in which matter particles
are treated as open strings propagating on a D3-brane, and the higher-dimensional
bulk space-time is punctured by D-particle defects. \underline{Lower}: Details of the process
whereby an open string state propagating on the D3-brane is captured by a D-particle
defect, which then recoils. This process involves an intermediate composite state that persists for a period
$\delta t \sim \alpha ' E$, where $E$ is the energy of the incident string state. This distorts the surrounding
space-time during the scattering process, leading to an effective refractive index,
but not birefringence. Components of the recoil velocity perpendicular to the D3-brane world lead to
vacuum energy contributions and thus target-space supersymmetry breaking.}%
\label{fig:recoil}%
\end{figure}

The model illustrated in the upper
panel of Fig.~\ref{fig:recoil}~\cite{Dfoam} will serve as our concrete D-brane approach to
the phenomenology of space-time foam.
In it, after appropriate compactification our Universe is represented  as a Dirichlet three-brane (D3-brane),
on which conventional particles propagate as open strings. This 3-brane
propagates in a 10-dimensional bulk
space-time containing orientifold planes, that is punctured by D-particle defects~%
\footnote{Since an isolated D-particle cannot exist~\cite{polchinski}
because of string gauge flux conservation, the presence of a D-brane is essential.}.
D-particles cross the D3-brane world as it moves
through the bulk. To an observer on the D3-brane, these crossings constitute a
realization of `space-time foam' with defects at space-time events due to
the D-particles traversing the D3-brane:
we term this structure `D-foam'. When the open strings encounter D-particles in
the foam, their interactions involve energy-momentum exchange that cause the
D-particles to recoil.

If there is no relative
motion of branes in the bulk, target-space supersymmetry,
implies the vanishing of the ground-state energy of the configuration.
However, if there is relative bulk motion, supersymmetry is broken and there are non-trivial forces among
the D-particles, as well as between the D-particles and the brane world and orientifolds~\cite{Dfoam}.
The resulting non-zero contribution to the energy is proportional to $v^2$ for \emph{transverse} relative
motions of branes with different dimensionalities, and to $v^4$ for branes of the same dimensionality~%
\footnote{There is \emph{no} contribution to the energy of a D-brane world from motions of other branes in
directions \emph{parallel} to its longitudinal directions.}. There is also a dependence on the relative
distances of the various branes. In particular, the interaction of a single D-particle that lies far away from a
D8-brane world, and moves adiabatically with a
small velocity $v_\perp$ in a direction transverse to the brane, results in the following potential~\cite{Dfoam}:
\begin{eqnarray}
\label{D0-D8-long}
\mathcal{V}_{D0-D8}^{long} =  + \frac{r\,(v^{\rm long}_\perp)^2}{8\pi\alpha^{\prime}} + \dots ~, ~r \gg \sqrt{\alpha^{\prime}}~,
\end{eqnarray}
where the $\dots$ indicate velocity-independent parts, that are cancelled by the orientifold planes in the model
of~\cite{Dfoam}, and do not play any r\^ole in our discussion below. On the other hand, a D-particle close to the
D8-brane, at a distance $r' \ll \sqrt{\alpha '} $ and  moving adiabatically in the perpendicular direction
with a velocity $v^{\rm short}_\perp$, induces the following potential:
\be\label{pot1}
\mathcal{V}_{D0-D8}^{short}=  - \frac{\pi\alpha^\prime (v_\perp^{\rm short})^2}{12{r'}^3} + \dots~, \quad r < \sqrt{\alpha '}~,
\ee
where again the $\dots$ denote velocity-independent terms that are cancelled in  the model of~\cite{Dfoam},
as mentioned previously.
This D-foam model involves different configurations of D-particles~\cite{Dfoam}, and one must average
over appropriate populations and quantum fluctuations~\cite{sarkar} of D-particles with bulk
(nine-dimensional) recoil velocities $v_i$.

\subsection{D-foam and the effective target space-time metric}

Recoil fluctuations {\it along} the D3-brane correspond to a stochastic fraction of the incident momentum of the open-string
particle~\cite{Dfoam,sarkar}. Due to the D-particle fluctuations in the direction transverse to the brane worlds, there
would also be contributions to the potential energy of the brane, (cf (\ref{D0-D8-long}), (\ref{pot1}), where now the
quantities involving $v^{\rm short, long}_\perp$ should be averaged over populations of D-particles and
quantum fluctuations). We can plausibly use a general parametrization of the foam fluctuations as follows:
\begin{eqnarray}\label{foam}
&& \delta v_A = g_s \frac{r_\parallel}{M_s} p_A~, A=1,2,3 \quad \Rightarrow \quad \langle r_\parallel \rangle = 0, ~ \langle r_\parallel^2 \rangle = \sigma_\parallel ^2 \ne 0~\nonumber \\
&& v^{\rm short}_\perp \equiv v^{\rm short}_\alpha~, \alpha =4, \dots 9  \quad \Rightarrow \quad \langle v^{\rm short}_\alpha  \rangle = 0, ~ \langle v^{\rm short}_\alpha v^{\rm short}_\beta  \rangle = \delta_{\alpha\beta} {\sigma_{\rm short}'}^2 \ne 0~; \alpha, \beta = 4 \dots 9~, \nonumber \\
&& v^{\rm long}_\perp \equiv v^{\rm long}_\alpha~, \alpha =4, \dots 9  \quad \Rightarrow \quad \langle v^{\rm long}_\alpha  \rangle = 0, ~ \langle v^{\rm long}_\alpha v^{\rm long}_\beta  \rangle = \delta_{\alpha\beta} {\sigma_{\rm long}'}^2 \ne 0~; \alpha, \beta = 4 \dots 9~,
\nonumber \\
&& \langle v^{\rm short,long}_\alpha v^{\rm short, long}_A  \rangle = 0~.
\end{eqnarray}
In general, stochastic foam implies vanishing correlators of odd powers of the recoil velocity $v_i$,
but non-trivial correlators of even powers. Above, $\langle \dots \rangle $ indicates averaging over both
quantum fluctuations and the ensemble of D-particles in the foam, and indices $A=1,2,3$ denote the longitudinal
dimensions of the D3-brane world. The  $\delta v_A$, $A=1,2,3$, represent the recoil velocity components of the
D-particle during scattering. In the absence of any interaction with matter strings there is also a velocity $v_i$
expressing the quantum fluctuations of individual D-particles.

In view of the opposite signs of the contributions (\ref{D0-D8-long}), (\ref{pot1}), 
in a stochastic foam situation it is possible for the effective potential on the brane world to vanish on average. However,
the spontaneous breaking of global supersymmetry is possible even in such a case, as we discuss below, and
also the breaking of local supersymmetry through the coupling to gravity, through the
special r\^ole played by the dilaton field explained above.

To explore this possibility, 
we consider in detail the interaction of an open-string particle with a heavy, non-relativistic D-particle
described as in~\cite{recoil} by the process of capture and release by the D-particle of a open string representing a 
neutral flavoured particle such as a neutrino, cf the lower figure in Fig.~\ref{fig:recoil}, which
may induce multi-fermion interactions of the form discussed in the previous Section.
As shown in~\cite{recoil}, the recoil excitation of a D-particle during its non-trivial scattering (capture) with an
open-string state propagating on the brane world as in Fig.~\ref{fig:recoil}) results in a distortion of the neighboring 
space-time which, for relatively long times after the scattering, is described by the following flat metric that
depends on the momentum transfer:
\begin{equation}\label{recoilmetric}
ds^2 = dt^2 - \delta_{ij} dx^i dx^j - 2 u_i dx^i dt
\end{equation}
where
\begin{equation}
u_i = g_s\frac{\Delta p_i}{M_s}
\label{recveldef}
 \end{equation}
is the D-particle recoil velocity and $\delta p_i$ the open-string-state momentum transfer during the scattering/capture process. 
This metric is Finsler-like, in that the recoil velocity $u_i$ depends in general on the open-string particle velocity.
The D-particles are assumed heavy, with their mass $M_/g_s \gg \Delta p_i$, so that $u_i \ll 1$ is non-relativistic.
This metric is nothing other than the induced metric, from the point of view of a passive observer, 
under a Gallilean transformation
$t \rightarrow t, \quad x^i \rightarrow x^i + \delta^{i}_{\, j}u_j t$. It is worth noticing that, 
from a world-sheet $\sigma$-model point of view,
which provides a first-quantized description of the dynamics of the composite string/recoiling-D-particle 
state following the scattering event
shown in the lower picture of Fig.~\ref{fig:recoil}, such a metric arises in the restoration of the world-sheet conformal 
symmetry that was disturbed by the recoil operators. This restoration was achieved by means of a Liouville dressing
of the $\sigma$ model, combined with the identification of target time with (the world-sheet zero mode of) the Liouville 
field, in a way consistent with the logarithmic conformal algebra of recoil~\cite{recoil}.

The metric (\ref{recoilmetric}) can be diagonalized by means of appropriate transformations. Equivalently, in a 
T-dual formalism, the propagation of strings in a recoiling D-particle background, when the string excitations 
interact with the D0-brane defect in the way described above, can be mapped to the problem of a string in 
a constant `electric field' background of the form~\cite{mavroreview}:
\begin{equation}\label{electricfield}
F_{0i} = u_i ~.
\end{equation}
The reader should notice that, in view of (\ref{recveldef}), the `electric' field background is actually in the 
phase space of the string state, as it depends on the relevant momentum transport.
Assuming for simplicity that the electric field is along the $x^1$ direction,  
(\ref{electricfield}) implies that the string state propagates in the diagonal metric
\begin{eqnarray}
ds^2= (1 - u_1^2)dt^2 - (1 - u_1^2)(dx^1)^2 - \sum_{i \ne 1} (dx^i)^2~.
\label{diagrecmetric}
\end{eqnarray}
As a result of quantum collisions in a stochastic space-time foam, 
the recoil velocity fluctuates along different directions.
This prompts us to consider the superposition of metrics (\ref{diagrecmetric}) at each space-time point, 
averaged over populations of D-particle defects. Such a stochastic superposition is denoted by 
$\ll \dots \gg = \frac{1}{d} \langle \dots \rangle_{\mathcal{D}}$, where $d$
is the number of spatial directions along which the recoil velocity vector fluctuates,
in our case we take $d=3$, and $\langle \dots \rangle_{\mathcal{D}}$ denotes averages 
over statistical ensembles of $D$-particle defects in the foam. Assuming homogeneous and isotropic foam backgrounds with
stochastic fluctuations of the form:
\begin{eqnarray}
\langle u_i u_j \rangle_{\mathcal{D}} = \sigma^2(t) \delta_{ij}~, \quad i, j =1,2,3~,
\label{stochflct}
\end{eqnarray}
we obtain for the average quantum foam induced metric at each space-time point
\begin{eqnarray}
\ll ds^2 \gg = ( 1 - \sigma^2(t) )dt^2 - (1 - \frac{\sigma^2(t)}{3}) \sum_{i=1}^3 (dx^i)^2
\label{avrecmetric}
\end{eqnarray}
After rescaling the time coordinate appropriately, we may map the homogeneous and isotropic metric into a
diagonal metric of cosmological type in a conformal time $\eta$:
\begin{equation}\label{finalrecmetric}
\ll ds^2 \gg = C(\eta) \left( d\eta^2 - \sum_{i} (dx^i)^2 \right)~, \quad C(\eta) \equiv 1 - \frac{\sigma^2(\eta)}{3}~,
\end{equation}
where $\sigma(\eta) = \sigma(t(\eta))$ denotes the time-dependent fluctuation function of the homogeneous and isotropic foam.
The time dependence reflects the possibility of a density of defects crossing our brane world in Fig.~\ref{fig:recoil}
that is not constant in cosmic time.

From a $\sigma$-model point of view, the metric (\ref{finalrecmetric}) may be viewed as the 
$\sigma$-model-frame metric. In this sense, the curvature kinetic term in the low-energy effective action, 
representing the dynamics of the problem, does not have the canonical Einstein form, 
but rather the Jordan-frame form
\begin{equation}\label{stringgrav}
S_{\rm grav} = M_P^2 \int d^4 x \, e\, e^{-2\varphi} R(g) + \dots
\end{equation}
where $\varphi$ is the four-dimensional dilaton field and the $\dots$ denote the rest of the terms in the 
low-energy effective action, including kinetic terms of the field $\varphi$. Appropriate compactification or 
other bulk effects are ignored for brevity, and $M_P$ denotes the four dimensional Planck mass. 
This is in general different from the D-particle mass $M_s/g_s$, where $M_s$ is the string mass scale, 
depending on the details of the compactification.

The physical metric~\cite{aben} is related to the $\sigma$-model Jordan-frame metric by the following 
transformation (we restrict our discussion to four space-time dimensions in what follows, as we consider
a D3 brane world):
\begin{equation}\label{physmetric}
g_{\mu\nu}^{\rm phys} = e^{-2\varphi} g_{\mu\nu} = e^{-2\varphi + {\rm ln}C(\eta)} \eta_{\mu\nu}
\end{equation}
Thus, we see that one may absorb the time-dependent effects of the D-foam induced 
scale factor $C(\eta)$ in a redefinition of the dilaton field.

\subsection{Cosmological-type backgrounds}

Let us first discuss a case in which the dilaton $\varphi$ is stabilized to a constant by means of some unspecified
potential - which is not known at present in string theory, given that it depends on string-loop effects over 
which there is no control in general.
In such a case, the conformally flat, homogeneous and isotropic physical space-time metric (\ref{physmetric})
with $\eta$ (time) dependence only, assumes a form familiar in cosmology, and the physical effect of the fluctuating 
D-particle foam on the brane world is to induce a cosmological background. The constancy of the 
dilaton $\varphi$ implies that in this simplified scenario the gravitational sector of the low-energy theory 
consists only of cosmological backgrounds, over which propagate the matter fields, represented in the microscopic picture 
by open strings on the D3 brane world.

Whether the background corresponds to an expanding universe depends on the bulk density of the  D-particles
in the model illustrated in Fig.~\ref{fig:recoil}. It is natural to expect that the stochastic fluctuations $\sigma(\eta)^2$
of the foam-averaged D-particle recoil velocities are proportional to the density of the foam, in the sense that the 
higher the density the stronger the average of the fluctuations. In such a case,  an expanding effective metric 
(\ref{physmetric}) on the brane Universe is obtained if the density of D-particles on the brane reduces as time elapses.

In the presence of string matter, then, the Einstein cosmological equations for the background (\ref{physmetric}) 
with constant dilaton $\varphi$ impose a consistency condition on the fluctuations $\sigma^2(\eta)$, i.e., the scale factor $C(\eta)$, 
depending whether the era is dominated by matter or dark energy. Lack of knowledge of the exact dilaton potential 
prevents a more explicit discussion of the corresponding low-energy action.

\subsection{D-foam-driven supersymmetry breaking and the dilatino Goldstino field}

However, more interesting for us is an alternative possibility that we now concentrate upon, namely
the possibility of an equilibrium, non-expanding Universe in the presence of D-foam effects.
Such a scenario is realized if the dilaton field is time-dependent, but its entire time dependence is of the form
\begin{equation}\label{dilC}
\varphi = +\frac{1}{2}\,{\rm ln}C(\eta)~.
\end{equation}
We then see from (\ref{physmetric}) that the physical metric is just flat Minkowski, and all the low-energy effects of the 
foam are absorbed by the dilaton. The gravitational sector of this theory consists of just a cosmological dilaton background,
and the corresponding field-theory equations of motion are given by just a flat-space equation of motion for the scalar field.
As already mentioned above, the potential of the dilaton in string theory is not known, in general, so we consider
the following generic equation of motion for the equilibrium dilaton in the presence of D-foam:
\begin{equation}\label{dilpot}
\Box \varphi + \dots = \frac{\delta V(\varphi)}{\delta \varphi} \equiv V^\prime (\varphi)~,
\end{equation}
where $\dots$ denote higher derivative terms in the low-energy string action, and $V(\varphi) $ is the dilaton potential, 
which receives in general contributions also from higher string loops. The matter terms in the effective 
action couple to the dilaton via exponential factors, in general:
\begin{equation}
V(\varphi) \ni \sum_{i={\rm matter~species}}\int d^4 x e^{\gamma_i \varphi}{\mathcal L}^{(i)}_{\rm matter} ~, \, \gamma^{(i)} = {\rm const}~,
\end{equation}
where the factors $\gamma_i$ are in general different for different matter species, and are determined by scale symmetry.
Such terms also contribute to the dilaton potential in (\ref{dilpot}), which must be known
precisely in order to see whether the expression (\ref{dilC})
is a consistent solution of the equations of motion.
Depending on the exact form of the dilaton potential, it is possible that there are constant, time-independent  solutions 
with $\sigma^2=$ constant that correspond to a minimum of the dilaton potential.

Such a solution could have important implications for the spontaneous breaking of target-space supersymmetry since,
depending on the form of the dilaton potential,
he F-term of the dilaton chiral superfield may acquire a 
vacuum expectation value that breaks global target-space supersymmetry. The dilaton is one of the two fields of the 
chiral Barbero-Immirzi superfield, the other being a pseudoscalar axion. The dilaton suffices for supersymmetry breaking,
and one may assume that the axion has a zero expectation value in the D-foam background.
However, this assumption is not binding, and more general D-foam backgrounds may be found with a non-trivial axion.
In our case, in which the constant dilaton 
configuration arises in the presence of an appropriate stochastic fluctuation of the D-foam recoil velocities, 
it is the space-time foam 
that is responsible for the breaking of target supersymmetry, inducing dynamical breaking of local supersymmetry (supergravity).
The way this is achieved in principle has been described in the previous sections, and will now be discussed
in the specific D-foam context.

Since this D-foam scenario differs from standard ways of breaking supersymmetry spontaneously in
that the above-described background
(\ref{finalrecmetric}) is induced by stringy matter excitations interacting with the D-foam, one can talk about a
supersymmetry \emph{obstruction }~\cite{witten} rather than breaking, in the sense that the true vacuum of the theory 
(characterised by the absence of excitations) may still be supersymmetric, but the effects of supersymmetry breaking 
can be seen in the excitation spectrum via, e.g., mass splittings between fermions and bosons.

Our discussion of a dynamically-generated gravitino mass (\ref{ginomass}) in an
effectively Minkowski flat space-time background applies intact in the case of D-foam, 
since the background of the physical metric (\ref{physmetric}) implied
by the specific dilaton background (\ref{dilC}) is indeed Minkowski. We recall that,
from a microscopic brane Universe point of view (see Fig.~\ref{fig:recoil}),
in general one may obtain (anti-)de Sitter-type constant vacuum energy contributions, as a result of the forces exerted on 
the three-brane Universe by bulk D-particles, (\ref{D0-D8-long}), (\ref{pot1}),
upon averaging over D-particle populations using (\ref{foam}). In such a case, the transverse distance $r$ may be replaced 
by a constant representing an average value, usually assumed to be of order of the string length $\sqrt{\alpha^\prime}$.

For simplicity, we assume that the foam recoil-velocity fluctuations of the short- and long-distance defects are of the same order:
\begin{equation}
\langle v_\alpha^{\rm short, \rm long} v_\beta^{\rm short, \rm long} \rangle = \delta_{\alpha\beta} \delta_{\rm short\, long}~{\sigma'}^2~,
\label{foam2}
\end{equation}
in which case the contributions from the foam to the vacuum energy density on the brane can be negative, i.e., of anti-de-Sitter type and
compatible with supersymmetry in the three-brane world, and of order ${\sigma'}^2 $, provided the effects of the bulk defect 
populations near the brane overcome those of the distant D-particles, which is a possible outcome of the bulk distribution of 
the D0-particles in the models. Note that in this picture, the
quantum fluctuations of the D-particles along the three brane, $\langle v^2_\parallel \rangle = \sigma^2 $ may in general 
be different from ${\sigma'}^2$.

In such a case, the significance of the dilaton background (\ref{dilC}), 
which enforces a Minkowski space-time form for the background physical metric, is analogous to the imposition of
the cancellation of the cosmological constant contributions (\ref{condition}) in~\cite{smith}.
In such a case, the supersymmetry-breaking scale associated with the v.e.v. of the dilaton has the value
\begin{equation}\label{vevphi}
\langle \varphi \rangle = +\frac{1}{2}{\rm ln}\left|1 - \frac{\sigma^2}{3} \right| ,
\end{equation}
when can be expressed in terms of the bulk fluctuations ${\sigma'}^2$ (\ref{foam2}), in order to enforce the cancellation of the 
effective cosmological constant on the brane.  This cancellation involves the 
the string-loop induced contribution to the dilaton potential, and cannot be derived from first principles,
since the dilaton potential in string theory is not known. In this respect, our situation is no different from
any other approach to string phenomenology.

The presence of a non-trivial dilaton potential in the effective low-energy supergravity theory on the brane world
prompts us to consider conformal supergravity models with more general non-trivial superpotentials than that  in~\cite{gates}
in order to explore the possible effects of the Barbero-Immirzi (dilaton-axion) superfield. 
One such model is the conformal supergravity model in the Jordan frame
of~\cite{ferrara2}, which in component form contains among others the following gravitational, 
dilaton and four-gravitino terms in the notation and conventions of~\cite{ferrara2}:
\begin{eqnarray}\label{confsugra}
e^{-1} \mathcal{L}  &=&  -\frac{1}{6} \tilde{\Phi} \left[R(e) - \overline{\psi}_\mu \mathcal{R}^\mu \right] - \frac{1}{6} (\partial_\mu \tilde{\Phi})
\left(\overline{\psi}_\alpha \gamma^\alpha \psi^\mu \right) -V  + \frac{1}{96} \tilde{\Phi} \left[ \overline{\psi}^\rho \gamma^\mu \psi^\nu \left(\overline{\psi}_\rho  \gamma_\mu \psi_\nu + 2 \overline{\psi}_\rho  \gamma_\nu \psi_\mu \right) + \dots  \right]~, \nonumber \\
 \mathcal{R}^\mu & \equiv & \gamma^{\mu\rho\sigma}\left(\partial_\rho + \frac{1}{4} \omega_\rho^{ab} (e) \gamma_{ab} + {\rm auxiliary~fields}\right) \psi_\sigma ~,
\end{eqnarray}
where the notation $\gamma^{\rho\mu\dots} $ denotes appropriately-antisymmetrised products of $\gamma$ matrices,
$e$ denotes the vierbein, $\omega_\mu  (e)$ is the torsion-free spin connection, and the four-gravitino terms arise from the 
torsion of the supergravity model~\footnote{The action (\ref{confsugra})  reduces, for $\Phi = 6/\kappa^2$ (in our conventions 
and normalisation), to the standard N=1 supergravity action - with the terms $\overline{\psi}_\mu \mathcal{R}^\mu $ reducing to 
the standard kinetic term for the gravitino (\ref{rseta}), using standard properties of the product of three $\gamma$ matrices.}. 
The $\dots$ denote other structures, including kinetic terms for the scalar $\tilde{\Phi}$ and four-fermion terms coupling the 
gauginos of the model with the gravitinos, as well as appropriate auxiliary fields, which are not of interest to us here. 
Complete expressions can be found in~\cite{ferrara2}, to which we refer the interested reader. In that work, there are 
contributions to the scalar superfield also from the supersymmetrization of matter in the Standard Model. In our case, we 
concentrate only on the gravitational (D-foam) contributions to the (dilaton) scalar field $\tilde{\Phi}$, which is
one of the scalar components of the Barbero-Immirzi superfield.

Comparing (\ref{confsugra}) with the string effective action in the so-called $\sigma$-model (or Jordan) frame, (\ref{stringgrav}), 
we obtain the following relation between the Barbero-Immirzi superfield and the dilaton field $\varphi$ in our normalisation (up to numerical factors that we do not write explicitly):
\begin{equation}\label{immdil}
-\tilde{\Phi}= e^{-2\varphi}
\end{equation}
with constraints to ensure the positivity of $-\tilde{\Phi}$.
One can pass to the Einstein frame (whose metric is denoted by a superscript $E$), 
in which the curvature terms in (\ref{confsugra}) are canonically normalized with coefficient $\frac{1}{\kappa^2}$,
by redefining the metric (in the normalisation of \cite{ferrara2}):
\begin{equation}
g^E_{\mu\nu} = \Omega^{2} g_{\mu\nu} ~, \quad \Omega^2 = - \frac{1}{3}\tilde{\Phi} > 0 ~,
\end{equation}
and then normalising the kinetic terms of the gravitino terms, by absorbing appropriate powers of $\tilde{\Phi}$ into the field $\psi_\mu$.
The Einstein-frame version of the action (\ref{confsugra}) then reads (using (\ref{immdil}) and re-instating units of the gravitational 
constant $\kappa^2=8\pi G_N $ for completeness):
 \begin{eqnarray}\label{confsugra2}
&& \mathcal{L}^E  (e^E)^{-1} =  -\frac{1}{2\kappa^2} R^E(e^E)  + \frac{1}{2} \epsilon^{\mu\nu\rho\sigma} \overline{\psi '}_\mu  \gamma_5 \gamma_\nu D^E_\rho {\psi'}_\sigma -  e^{2\varphi}\,V^E
 + \nonumber \\ &~& \frac{11 \kappa^2}{16} e^{-2\varphi} \left[ (\overline{\psi'}_\mu  {\psi'}^\mu )^2  - (\overline{\psi'}_\mu  \gamma_5 {\psi'}^\mu )^2 \right]
+ \frac{33}{64} \kappa^2 e^{-2\varphi} \, \left(\overline{\psi'}^\rho \gamma_5  \gamma_\mu {\psi'}_\rho  \right)^2  + \dots   =
\nonumber \\
 && -\frac{1}{2\kappa^2} R^E(e^E)  + \frac{1}{2} \epsilon^{\mu\nu\rho\sigma} \overline{\psi '}_\mu  \gamma_5 \gamma_\nu D^E_\rho {\psi'}_\sigma -  e^{2\varphi}\,V^E
 +  \nonumber \\ &~&   \rho^2(x)  +  \frac{\sqrt{11}}{2}\kappa \rho (x) e^{-\varphi}\, \left(\overline{\psi'}_\mu {\psi'}^\mu \right)  + \pi^2(x)  +  \frac{\sqrt{11}}{2}e^{-\varphi}\, \kappa \, i \pi (x)  \left(\overline{\psi'}_\mu \gamma_5 {\psi'}^\mu \right) + \nonumber \\ &~& \frac{\sqrt{33}}{2} \kappa \, e^{-\varphi}\, i \lambda^\nu \left(\overline{\psi'}^\rho \gamma_5  \gamma_\nu {\psi'}_\rho \right) + \dots ~,
 \end{eqnarray}
where $\psi'_\mu$ denotes the canonically-normalised gravitino with standard kinetic term as in $N=1$ supergravity,
and the $\dots$ denote structures, including auxiliary fields, that are not of interest here. In writing
(\ref{confsugra2}) we have expanded the four-gravitino terms into detailed structures to exhibit explicitly the terms that
generate masses, and we linearise the four-gravitino terms. The condensate of interested to us is the v.e.v. of the 
linearizing field $\rho (x)$.

The reader should notice that the coefficients of the gravitino-$\rho$ interaction terms in (\ref{confsugra2}) are now modified 
compared to (\ref{effactionlinear}) by dilaton-dependent factors $\sim e^{-\varphi}$, 
being proportional not to $\kappa$ but to:
\begin{equation}\label{kaptilde}
{\tilde \kappa} \equiv \kappa e^{-\varphi}~. 
\end{equation}
Using (\ref{vevphi}), we observe that the coefficients of the four-gravitino terms in (\ref{confsugra2}) are of order
\begin{equation}
\frac{11 \, \kappa^2}{16} \left(1 -\frac{\sigma^2}{3}\right)^{-1} ~.
\end{equation}
We observe that the foam fluctuations tend to increase the interaction as compared with the Minkowski background case~\cite{smith}.
In the weak foam backgrounds we have considered in our work so far, the form of the induced metric deformation 
(\ref{finalrecmetric}) implies that $\sigma^2 \ll 1$, otherwise the signature of the space and time components of the 
induced metric are reversed.
This situation implies that the foam fluctuations are sufficiently weak that the gravitational fluctuations of the background cannot be
ignored, which, according to the arguments of \cite{odintsov} and our previous discussion, destabilizes the  putative gravitino 
condensate.

However, one may consider the embedding of the D-foam model~\cite{Dfoam} in a cosmological framework~\cite{Dfoamcosmo}, in which
case strong negative dilaton backgrounds in the conformal supergravity framework are also allowed, which include not only
the foam contributions (\ref{vevphi}) but also much stronger cosmological time dependences. In particular, for late times in the 
history of the Universe, far away from the  inflationary phase, one may have $\varphi \ll -1$,  
in which case $e^{-2\varphi} \gg 1$. One such background is the cosmological run-away linear-dilaton case~\cite{aben}, 
in which at asymptotically late cosmic Robertson-Walker time $t$ in the history of the Universe,
\begin{equation}\label{runaway}
\varphi \sim -{\rm ln}~t \to -\infty ~.
\end{equation}
Such backgrounds also arise in our non-equilibrium Liouville string approach to quantum gravity~\cite{liouv,Dfoam,Dfoamcosmo}, 
in which the world-sheet zero mode of the Liouville dressing field plays the
r\^ole of time. When such backgrounds are considered in conformal supergravity (\ref{confsugra2}), 
which is a natural effective low-energy framework for string theory, the above considerations imply the
presence of strong coupling four-gravitino interactions that could lead to stable condensates, 
unaffected by weak gravitational fluctuations.

This D-foam model provides a specific realization of the scenario for
global and hence local supersymmetry breaking via the v.e.v. of a dilaton $F$ term
that we proposed in the previous Section, which is based on
D-particle stochastic recoil velocity fluctuations ${\sigma '}^2$ perpendicular to the brane Universe, cf, Fig.~\ref{fig:recoil}
and (\ref{foam}). As we have seen (\ref{pot1}) the contributions of forces exerted on our brane world by
neighbouring bulk D-particles are negative and of anti-de-Sitter type, provided the density of
such defects is more or less constant for late eras. The latter can be arranged by a suitable bulk density of 
D-particles in the foam~\cite{Dfoamcosmo}, that is consistent with the observed cosmology at  late eras.
In fact, using (\ref{pot1}) and averaging over populations of nearby bulk D-particles we obtain a
vacuum energy on the brane world of the form of a cosmological constant proportional to ${\sigma'}^2 = \ll v_\perp^2 \gg $, 
where $v_\perp $ denotes the velocity of the D-particles transverse to the brane world.  The reader should recall that  
any D-particle movement parallel to the brane world does not yield any contributions to the brane vacuum energy in our model.

The D-foam recoil velocity bulk fluctuations $\sigma^2, \, {\sigma'}^2$ are free parameters of the model at the
present stage of our understanding. The quantum fluctuations of the D-particles parallel to the brane 
Universe are associated with the momentum transfer and hence influenced by the momentum of the matter excitation, 
whereas those perpendicular to it are related to properties of the foam \emph{per se}.
It is natural then  to assume that in our low-energy effective supersymmetric field theory framework
the induced supersymmetry breaking scale $f$ is of order of the foam fluctuations ${\sigma'}^2$:
\begin{equation}\label{susyscale}
f^2 = O({\sigma'}^2) M_P^4 ~,
\end{equation}
where $M_P^2 = V^c M_s^2 $, with $M_s$ the string scale and $V^c$ the volume of the
compact extra dimensions in units of the string length $\sqrt{\alpha'} =1/M_s$.
In this way, for sufficiently small D-particle recoil fluctuations ${\sigma'}^2 \ll 1$ one may obtain
supersymmetry breaking scales that are low compared to the four-dimensional Planck mass, 
in which case the infrared limit used in our analysis in this paper is applicable.

In this picture, the supersymmetry of the vacuum is broken spontaneously, in the sense that the 
recoil velocity fluctuations that affect the v.e.v. of the $F$-term of the dilaton field are vacuum properties, 
and supersymmetry transformations are formally invariances of the effective Lagrangian.
This implies that the supersymmetric partner of the dilaton is the Goldstino field, for rthe easons discussed above.
The r\^ole of the initially negative (anti-de-Sitter) cosmological constant of the Volkov-Akulov field is then played 
in this  framework by the vacuum energy (\ref{pot1}) induced by nearby bulk D-particles~\cite{Dfoam}.

As explained above, the coupling of the dilatino with the rest of the low-energy conformal supergravity is 
described in the infrared by the Volkov-Akulov Lagrangian (\ref{va2}), (\ref{goldstino}), (\ref{va3}),
which realizes global supersymmetry non-linearly, coupling to the Lagrangian (\ref{confsugra2})
via the goldstino/gravitino coupling of (\ref{va2b}). The Goldstino field is then eaten by the gravitino, 
which becomes massive via the formation of the condensate, as discussed in the previous Section. 
In the specific D-foam model discussed in this Section,
the condensate $\rho $ contributes a positive term ($\rho^2$) to the brane vacuum energy
able to cancel the anti-de-Sitter type contributions due to $-f^2$  (\ref{susyscale}) in (\ref{goldstino}).  
In our low-energy string-induced conformal supergravity framework, the full effective 
potential is complicated, since the precise form of the dilaton potential is not known in string theory.
Hence, the cancellation of the effective cosmological constant on the brane is an issue that is far from being resolved here.

Nevertheless, we now offer plausibility arguments how such a cancellation may be achieved in our context. 
The key point to realize is that, once the effective conformal supergravity action is considered in the run-away dilaton limit 
(\ref{runaway}), the effective coupling of the four-gravitino terms is no longer given by the gravitational constant, 
but by the much stronger effective coupling (\ref{kaptilde}). In practice we may consider the runaway dilaton limit as 
describing a long but finite time after, say, inflation, at which $\varphi \ll -1$ 
but finite: the universe may enter a phase in which the dilaton is frozen at a large but finite value. 

\subsection{Analysis of One-Loop Effective Potential}

In this limit the effective action (\ref{confsugra2}) can lead to an effective potential that, at one-loop order,
is obtained by following exactly the analysis of~\cite{smith2}, but replacing the gravitational coupling in 
that work by the new coupling $\tilde{\kappa}$ (\ref{kaptilde}). 
The resulting effective potential for the field $\rho(x)$ at one-loop order is then found to be~\cite{smith2}:
\begin{eqnarray}\label{effpotsugra}
e^{-1} \, \mathcal{V}_{\rm eff} & = & \frac{4}{(2\pi)^4} \int^\Lambda  d^4 p
{\rm ln}\left(1 + 11 \frac{{\tilde \kappa}^2 \rho^2}{p^2} \right) + f^2 - \rho^2 \nonumber \\
& = &
(\frac{1}{4\pi^3})\{ \frac{121}{2} {\tilde \kappa}^4 \rho^4 \left[
{\rm ln}(11 {\tilde \kappa}^2 \rho^2 /\Lambda^2) - \frac{1}{2} \right]
+ 11 {\tilde \kappa}^2 \rho^2 \Lambda^2 \} + f^2 - \rho^2 ~,
\end{eqnarray}
where $\Lambda$ is cut-off used to regularize the UV 
divergences of the non-renormalizable effective low-energy $N=1$ supergravity theory. 
Following~\cite{smith2}, one observes that for the values 
\begin{equation}\label{momcutoff}
\Lambda = {\tilde \kappa}^{-1}~, \quad \frac{1}{f} = 1.796 {\tilde \kappa}^2
\end{equation}
the effective potential is non-negative and has a minimum at zero, therefore leading to a vanishing 
cosmological constant on our brane world, consistent with a Minkowski background, allowing 
us to interpret terms of the form $<\rho> {\overline \psi}'_\mu {\psi'}^\mu $ as corresponding to gravitino mass terms.

\begin{figure}[ht]
\centering
\includegraphics[width=9cm]{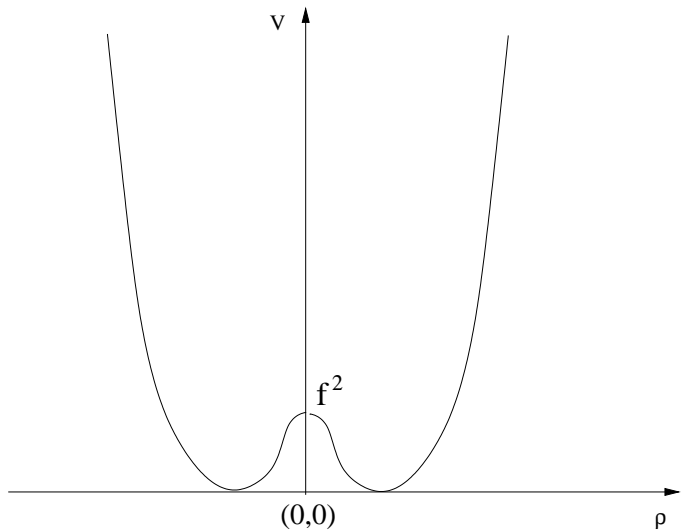} 
\caption{\it The one-loop effective potential for the linearizing field $\rho(x)$ in the effective action (\ref{confsugra2}), 
which is suitable for dynamical breaking of local supersymmetry and the generation of a gravitino mass. 
For certain values of the relevant parameters the potential has a minimum at zero, indicating an effective vanishing 
cosmological constant.}%
\label{fig:effpot}%
\end{figure}

The numerical analysis of~\cite{smith2} shows that the effective potential for the $\rho$ field acquires a minimum at 
\begin{equation}
\rho_{\rm min} = <\rho> = \pm 0.726 
\end{equation}
at which it vanishes, cf Fig.~\ref{fig:effpot}. The gravitino mass term in (\ref{confsugra2}) then takes 
the following form, cf (\ref{identity}):
\begin{equation}
-m_{3/s} {\overline \psi'}_\mu \Gamma^{\mu\nu} {\psi'}_\nu = 
-\frac{1}{2} m_{3/2} {\overline \psi'}_\mu {\psi'}^\mu ,
\end{equation}
with 
\begin{equation}\label{gravinomassfinal}
m_{3/2} = \sqrt{11} {\tilde \kappa}^{-1} \rho_{\rm min} = 2. 408 {\tilde \kappa}^{-1} = 3.227 \, \mathcal{O}({\sigma'}^2) M_P \ll M_P
\end{equation}
for a small foam contribution, where we used (\ref{susyscale}) and (\ref{momcutoff}). 

We recall that cosmological and astrophysical constraints are compatible either with scenarios where the
gravitino is the lightest supersymmetric particle and makes a contribution to dark matter, or with scenarios
where the gravitino weighs a few TeV or more, in which case it is unstable and decays sufficiently early
not to upset the agreement between standard Big-Bang nucleosynthesis and astrophysical observations
of light-element abundances. Indeed, the decays of a gravitino in this mass range might even improve
agreement with the measured abundance of $^7$Li~\cite{CEFLOS}.

The important physical difference of our case from that of~\cite{smith2} lies in the fact that, as a result of the dilaton factors, 
the effective coupling ${\tilde \kappa}$ appearing in (\ref{momcutoff}) is much larger than the gravitational constant 
$\kappa = 1/M_P$.  Because of this enhancement of the four-gravitino interactions in (\ref{confsugra2})
by dilaton factors $e^{-2\varphi}$ (\ref{kaptilde}),  one can argue that metric fluctuations around the space-time foam 
background cannot destabilize the condensate, and thus the scenario presented here implements the original arguments 
of~\cite{smith}, avoiding the objections of~\cite{odintsov}. Indeed,
for sufficiently low supersymmetry-breaking scales (\ref{susyscale}), the relevant excitations are those with 
momenta below $\sqrt{f} \ll M_P$, where
quantum-gravity fluctuations are irrelevant.

Before closing this section we remark on another issue associated with a potentially non-trivial r\^ole of the
Immirzi-Nieh-Yan topological density (\ref{nysugra}) in the supergravity case, which
involves the total derivative of the gravitino current.
An on-shell solution of the gravitino field equations obtained from the appropriate Lagrangian in a recoiling 
space-time foam background may contain singularities, which could make such total derivative contributions 
non-trivial, in which case the Barbero-Immirzi parameter appearing in the respective Holst actions could
play a physical r\^ole. If such were the case, there would be a direct analogy of the Barbero-Immirzi 
parameter with the $\theta$ parameter of QCD in the presence of colour magnetic monopoles.

In our recoiling D-particle background, the  space-time metric deformation induced by the recoil of a single D-particle 
during its interaction with an open string excitation
includes off-diagonal terms of the following form (before averaging over D-particle populations in the foam)~\cite{recoil}:
\begin{equation}\label{recmetr}
 g_{0i} = u_i X^0 \Theta (X^0) ,
\end{equation}
where $X^0 $ is the time. For long times after the impact with the defect, the induced metric takes the form 
(\ref{recoilmetric}), which is then averaged over D-particle populations in the foam to produce the situation 
we have analysed in this work.
However, close to the moment of impact of the string state on the D-particle defect, and its splitting as in Fig.~\ref{fig:recoil},
the space-time structure is that of a Rindler wedge~\cite{winstan}. There are singularities
of the induced metric Christoffel connection in such space-time backgrounds that have a $\delta (X^0)$ structure.
Such singularities are integrable, though, and hence they do not yield any non-trivial contributions to the 
Holst term (\ref{nysugra}).
Nevertheless, the effective field theory close to the moment of impact may be quite different from the effective $N=1$ 
supergravities we have discussed in this work, which are relevant for situations far away from that moment. 
In fact, it was argued in~\cite{winstan} that a metric of the form
(\ref{recmetr}) arises as a consistent solution of the gravitational part of an effective $(d+1)$-dimensional 
action on a $d$-dimensional brane world,  involving a Gauss-Bonnet curvature-squared combination coupled to a 
dilaton field,
$\int d^{d+1} x e^{-2\varphi} \left(R_{\mu\nu\rho\sigma}^2 - 4 R_{\mu\nu}^2 + R^2 \right) + \dots $.
Supersymmetrising such terms would lead to supergavities with higher-order curvature terms, 
of the form considered in low-energy actions of superstrings and in a field-theory framework in \cite{ferrara3}. 
These actions are not the subject of our present discussion, but are viewed as describing the short-distance properties 
of the foam, while the considerations in our work, based on conformal $N=1$ supergravities 
with conventional Einstein curvature terms, describe the long-distance (infrared) properties of the D-foam. 
The issue whether other singular backgrounds could also produce such 
contributions and break supersymmetry is an open one that we do not discuss further here.

\section{Conclusions}

In this paper we have highlighted the r\^ole in global and local supersymmetry breaking that might be
played by the Barbero-Immirzi parameter in an effective low-energy supergravity theory valid in the
infrared limit. We have stressed the potential analogies with the the $\theta$ vacuum parameter in a
non-Abelian gauge theory. In that case non-trivial topological configurations are thought to play an
important r\^ole in the formation of a chiral symmetry-breaking condensate. We have argued that
non-perturbative events might play an analogous r\^ole in a suitably compactified string theory.
Specifically, we have pointed out that D-particles moving through the bulk appear as space-time events
when crossing the D-brane Universe, and have shown that interactions with open-string particle 
excitations that are associated with such events could give rise to four-fermion interactions. 
Combined with a suitable string contribution to the effective potential of the dilaton component of the
the Barbero-Immirzi superfield multiplet, this mechanism may lead to condensation with vanishing 
net vacuum energy in the brane Universe that breaks supersymmetry both globally and locally.
In this scenario, the dilatino component of the Barbero-Immirzi supermultiplet is represented by a
non-linear effective Lagrangian of the Volkov-Akulov type, and is eaten by the gravitino, giving it a
mass.

In closing, we emphasize that the general features of this Barbero-Immirzi mechanism for breaking 
supersymmetry may have wider validity than the specific D-particle recoil implementation that we have discussed here.
Specifically, in the framework of conformal supergravity~\cite{ferrara2} in the Einstein frame (\ref{confsugra2}), all we need 
for our arguments on the dynamical breaking of local supersymmetry through the formation of condensates is some dynamical mechanism yielding a v.e.v. of the dilaton field of appropriate size, so that the induced effective coupling of the 
four-gravitino interactions becomes stronger than the gravitational coupling, overcoming its destabilizing effects.

In our D-foam framework we have used the supersymmetric dilaton/axion multiplet identified with the complex
Barbero-Immirzi superfield $\Phi$ to provide, in the infrared, the Goldstino superfield associated with the breaking of 
supersymmetry. However, one may consider in addition more general scalar fields, e.g., Standard Model scalar 
multiplets such as Higgs fields, as in the analysis of \cite{ferrara2}. Such fields have, in the corresponding Jordan frame (or 
$\sigma$-model frame in the string picture of~\cite{aben}), non-minimal couplings with the curvature that might lead to acceptable 
slow-roll conditions for inflation in the early Universe~\cite{shaposhnikov, higgsdil, ferrara2}. Connecting such 
non-minimally coupled scalar multiplets to gravitino condensates, as discussed here, might provide a link to minimal 
inflation as suggested in~\cite{alvarez}. However, this issue is a delicate one, depending on the details of the scalar field 
potential, which in our string dilaton case is not known. {\it Une affaire \`a suivre.}

\section*{Acknowledgements}

This work was supported in part by the London Centre for
Terauniverse Studies (LCTS), using funding from the European Research
Council via the Advanced Investigator Grant 267352.


\begin{thebibliography}{99}

\bibitem{ashtekar} A. Ashtekar, Phys. Rev. Lett. \textbf{57}, 2244 (1986); Phys. Rev. D \textbf{36}, 1587 (1987).


\bibitem{asht2} A. Ashtekar, Contemp. Math. \textbf{71}, 39  (1988).

\bibitem{barbero} F. Barbero, Phys. Rev. D \textbf{51 }, 5498 (1995); \emph{ibid.} \textbf{51 }, 5507 (1995).

\bibitem{loopqg} A. Ashtekar, C. Rovelli and L. Smolin, Phys. Rev. Lett. \textbf{69 }, 237 (1992);
A. Ashtekar and J. Lewandowski, Class. and Quant. Grav. \textbf{21}, R53 (2004);
for a detailed review see: C. Rovelli, \emph{Quantum Gravity} (Cambridge Univ. Press, UK 2006), and references therein.

\bibitem{immirzi}
  G.~Immirzi,
  Class.\ Quant.\ Grav.\  {\bf 14}, L177-L181 (1997).
  [gr-qc/9612030].


\bibitem{holst}
  S.~Holst,
  Phys.\ Rev.\  {\bf D53}, 5966-5969 (1996).
  [gr-qc/9511026].

\bibitem{rovelli}
  A.~Perez and C.~Rovelli,
  Phys.\ Rev.\  D {\bf 73}, 044013 (2006)
  [arXiv:gr-qc/0505081];
L.~Freidel, D.~Minic and T.~Takeuchi,
  Phys.\ Rev.\  D {\bf 72}, 104002 (2005)
  [arXiv:hep-th/0507253].

\bibitem{mukho} B.~Mukhopadhyay,
  Class.\ Quant.\ Grav.\  {\bf 24}, 1433-1442 (2007).
  [gr-qc/0702062 [GR-QC]];
M.~Sinha, B.~Mukhopadhyay,
  Phys.\ Rev.\  {\bf D77}, 025003 (2008).
  [arXiv:0704.2593 [hep-ph]]. These authors considered  purely gravitational connections, with no torsion,
 but metric backgrounds that broke the axial symmetry. Nonetheless, by adding torsion in their formalism one recovers the
  four fermion terms induced by the Immirzi parameter.


\bibitem{neubert} F.~Giacosa, R.~Hofmann and M.~Neubert,
  JHEP {\bf 0802}, 077 (2008)
  [arXiv:0801.0197 [hep-th]];


\bibitem{biswas} S.~Alexander, T.~Biswas and G.~Calcagni,
  Phys.\ Rev.\  D {\bf 81}, 043511 (2010)
  [Erratum-ibid.\  D {\bf 81}, 069902 (2010)]
  [arXiv:0906.5161 [astro-ph.CO]];
S.~H.~S.~Alexander,
  arXiv:0911.5156 [hep-ph].

\bibitem{art}  A.~Ashtekar, J.~D.~Romano, R.~S.~Tate,
  Phys.\ Rev.\  {\bf D40}, 2572 (1989).



\bibitem{mercuri} S.~Mercuri,
  Phys.\ Rev.\  {\bf D73}, 084016 (2006).
  [gr-qc/0601013]; \emph{ibid.}
D {\bf 77}, 024036 (2008)
  [arXiv:0708.0037 [gr-qc]].



\bibitem{mercuri2} S.~Mercuri,
  arXiv:0903.2270 [gr-qc].
S.~Mercuri, V.~Taveras,
  Phys.\ Rev.\  {\bf D80}, 104007 (2009).
  [arXiv:0903.4407 [gr-qc]];
S.~Mercuri and A.~Randono,
  Class.\ Quant.\ Grav.\  {\bf 28}, 025001 (2011)
  [arXiv:1005.1291 [hep-th]]

\bibitem{alexandrov} S.~Alexandrov,
  Class.\ Quant.\ Grav.\  {\bf 25}, 145012 (2008)
  [arXiv:0802.1221 [gr-qc]].



\bibitem{ny} H. T. Nieh and M.L. Yan, J. Math. Phys. \textbf{23}, 373 (1982).


\bibitem{sengupta} G.~Date, R.~K.~Kaul, S.~Sengupta,
  Phys.\ Rev.\  {\bf D79}, 044008 (2009).
  [arXiv:0811.4496 [gr-qc]]; 
R.~K.~Kaul, S.~Sengupta,
  [arXiv:1106.3027 [gr-qc]]
  
  \bibitem{seng} S.~Sengupta,
  Class.\ Quant.\ Grav.\  {\bf 27}, 145008 (2010).
  [arXiv:0911.0593 [gr-qc]].


\bibitem{bifield} V.~Taveras and N.~Yunes,
  Phys.\ Rev.\  D {\bf 78}, 064070 (2008)
  [arXiv:0807.2652 [gr-qc]];

\bibitem{bifield2} G.~Calcagni and S.~Mercuri,
  Phys.\ Rev.\  D {\bf 79}, 084004 (2009)
  [arXiv:0902.0957 [gr-qc]].

\bibitem{ferrara} D.~Z.~Freedman, P.~van Nieuwenhuizen, S.~Ferrara,
  Phys.\ Rev.\  {\bf D13}, 3214 (1976);
D.~Z.~Freedman, P.~van Nieuwenhuizen,
  Phys.\ Rev.\  {\bf D14}, 912 (1976);
S.~Deser, B.~Zumino,
  Phys.\ Lett.\  {\bf B62}, 335 (1976).



\bibitem{tsuda} M.~Tsuda,
  Phys.\ Rev.\  {\bf D61}, 024025 (2000).
  [gr-qc/9906057].




\bibitem{kaul} R.~K.~Kaul,
  Phys.\ Rev.\  {\bf D77}, 045030 (2008).
  [arXiv:0711.4674 [gr-qc]].


\bibitem{kaulseng} S.~Sengupta, R.~K.~Kaul,
  Phys.\ Rev.\  {\bf D81}, 024024 (2010).
  [arXiv:0909.4850 [hep-th]].
  
  
\bibitem{gates} S.~J.~J.~Gates, S.~V.~Ketov and N.~Yunes,
  Phys.\ Rev.\  D {\bf 80}, 065003 (2009)
  [arXiv:0906.4978 [hep-th]].


\bibitem{gatesbook} See, for instance: S.~J.~Gates, M.~T.~Grisaru, M.~Rocek and W.~Siegel,
  Front.\ Phys.\  {\bf 58}, 1 (1983)
  [arXiv:hep-th/0108200].
\bibitem{komar} Z.~Komargodski and N.~Seiberg,
  JHEP {\bf 0909}, 066 (2009)
  [arXiv:0907.2441 [hep-th]].


\bibitem{va} D.~V.~Volkov and V.~P.~Akulov,
  Phys.\ Lett.\  B {\bf 46}, 109 (1973);
  Theor.\ Math.\ Phys.\  {\bf 18}, 28 (1974)
  [Teor.\ Mat.\ Fiz.\  {\bf 18}, 39 (1974)].


\bibitem{komar2} Z.~Komargodski and N.~Seiberg,
  JHEP {\bf 1007}, 017 (2010)
  [arXiv:1002.2228 [hep-th]].

\bibitem{arkani} N.~Arkani-Hamed, M.~Dine, S.~P.~Martin,
  Phys.\ Lett.\  {\bf B431}, 329-338 (1998).
  [hep-ph/9803432].





\bibitem{deser} S.~Deser, B.~Zumino,
  Phys.\ Rev.\ Lett.\  {\bf 38}, 1433 (1977).




\bibitem{smith}  R.~S.~Jasinschi, A.~W.~Smith,
  Phys.\ Lett.\  {\bf B173}, 297 (1986).

\bibitem{smith2} R.~S.~Jasinschi, A.~W.~Smith,
  Phys.\ Lett.\  {\bf B174}, 183 (1986).

\bibitem{nambu} Y.~Nambu and G.~Jona-Lasinio,
  Phys.\ Rev.\  {\bf 122}, 345 (1961);
  \emph{ibid}.  {\bf 124}, 246 (1961).


\bibitem{Dfoam} J.~R.~Ellis, N.~E.~Mavromatos and M.~Westmuckett,
Phys.\ Rev.\ D \textbf{70}, 044036 (2004) [arXiv:gr-qc/0405066];
\emph{ibid.} {\bf 71}, 106006 (2005)~.


\bibitem{candelas} P.~Candelas and D.~J.~Raine,
  Phys.\ Rev.\  D {\bf 12}, 965 (1975).




 \bibitem{odintsov} I.~L.~Buchbinder and S.~D.~Odintsov,
  Class.\ Quant.\ Grav.\  {\bf 6}, 1955 (1989).

\bibitem{fradkin} E.~S.~Fradkin and A.~A.~Tseytlin,
  Nucl.\ Phys.\  B {\bf 234}, 472 (1984).



\bibitem{odintsov2}   S.~D.~Odintsov,
  Phys.\ Lett.\  B {\bf 213}, 7 (1988).
see also
  Fortsch.\ Phys.\  {\bf 38}, 371-391 (1990).

\bibitem{polchinski} See for instance: J. Polchinski, \emph{String Theory},
Vol. \textbf{2} (Cambridge University Press, 1998).

\bibitem{sarkar} N.~E.~Mavromatos and S.~Sarkar,
  Phys.\ Rev.\  D {\bf 72}, 065016 (2005)
  [arXiv:hep-th/0506242].

\bibitem{recoil}  I.~I.~Kogan, N.~E.~Mavromatos and J.~F.~Wheater,
  Phys.\ Lett.\  B {\bf 387}, 483 (1996)
  [arXiv:hep-th/9606102];
J.~R.~Ellis, N.~E.~Mavromatos and D.~V.~Nanopoulos,
  Int.\ J.\ Mod.\ Phys.\  A {\bf 13}, 1059 (1998)
  [arXiv:hep-th/9609238].


\bibitem{mavroreview} N.~E.~Mavromatos,
  Found.\ Phys.\  {\bf 40}, 917-960 (2010).
  [arXiv:0906.2712 [hep-th]].




\bibitem{aben} I.~Antoniadis, C.~Bachas, J.~R.~Ellis, D.~V.~Nanopoulos,
  Phys.\ Lett.\  {\bf B211}, 393 (1988);
  Nucl.\ Phys.\  {\bf B328}, 117-139 (1989);
  Phys.\ Lett.\  {\bf B257}, 278-284 (1991);



\bibitem{witten}  E.~Witten,
  Int.\ J.\ Mod.\ Phys.\  {\bf A10}, 1247-1248 (1995).
  [hep-th/9409111].
  Mod.\ Phys.\ Lett.\  {\bf A10}, 2153-2156 (1995).
  [hep-th/9506101].
  [hep-th/9507121].
  In our context see: E.~Gravanis, N.~E.~Mavromatos,
  Phys.\ Lett.\  {\bf B547}, 117-127 (2002).
  [hep-th/0205298].
  J.~R.~Ellis, N.~E.~Mavromatos, D.~V.~Nanopoulos, M.~Westmuckett,
  Int.\ J.\ Mod.\ Phys.\  {\bf A21}, 1379-1444 (2006).
  [gr-qc/0508105].

  \bibitem{ferrara2} S.~Ferrara, R.~Kallosh, A.~Linde, A.~Marrani, A.~Van Proeyen,
  Phys.\ Rev.\  {\bf D82}, 045003 (2010).
  [arXiv:1004.0712 [hep-th]].

  \bibitem{Dfoamcosmo} J.~R.~Ellis, N.~E.~Mavromatos, D.~V.~Nanopoulos, M.~Westmuckett,
  Int.\ J.\ Mod.\ Phys.\  {\bf A21}, 1379-1444 (2006).
  [gr-qc/0508105];
  Int.\ J.\ Mod.\ Phys.\  {\bf A26}, 2243-2262 (2011).
  [arXiv:0912.3428 [astro-ph.CO]].


  \bibitem{liouv} J.~R.~Ellis, N.~E.~Mavromatos, D.~V.~Nanopoulos,
  Chaos Solitons Fractals {\bf 10}, 345-363 (1999).
  [hep-th/9805120] and references therein.

 \bibitem{winstan} J.~R.~Ellis, P.~Kanti, N.~E.~Mavromatos, D.~V.~Nanopoulos, E.~Winstanley,
  Mod.\ Phys.\ Lett.\  {\bf A13}, 303-320 (1998).
  [hep-th/9711163].

 \bibitem{ferrara3} S.~Ferrara, S.~Sabharwal, M.~Villasante,
  Phys.\ Lett.\  {\bf B205}, 302 (1988).
  see also: S.~Ferrara, P.~Fre, M.~Porrati,
  Annals Phys.\  {\bf 175}, 112 (1987) and references therein.
  
\bibitem{CEFLOS}
R.~H.~Cyburt, J.~Ellis, B.~D.~Fields, F.~Luo, K.~A.~Olive and V.~C.~Spanos,
  JCAP {\bf 1010} (2010) 032
  [arXiv:1007.4173 [astro-ph.CO]].

\bibitem{shaposhnikov} F.~L.~Bezrukov, M.~Shaposhnikov,
  Phys.\ Lett.\  {\bf B659}, 703-706 (2008).
  [arXiv:0710.3755 [hep-th]].
  JHEP {\bf 0907}, 089 (2009).
  [arXiv:0904.1537 [hep-ph]]
F.~Bezrukov, A.~Magnin, M.~Shaposhnikov, S.~Sibiryakov,
  JHEP {\bf 1101}, 016 (2011).
    [arXiv:1008.5157 [hep-ph]].
  
\bibitem{higgsdil}  J.~Garcia-Bellido, J.~Rubio, M.~Shaposhnikov, D.~Zenhausern,
  [arXiv:1107.2163 [hep-ph]].
  
  \bibitem{alvarez} L.~Alvarez-Gaume, C.~Gomez, R.~Jimenez,
  Phys.\ Lett.\  {\bf B690}, 68-72 (2010).
  [arXiv:1001.0010 [hep-th]];
  JCAP {\bf 1103}, 027 (2011).
  [arXiv:1101.4948 [hep-th]].
\end{thebibliography}
\end{document}